\documentclass{article} 
\usepackage{iclr2026_conference,times}

\usepackage[T1]{fontenc}
\usepackage[utf8]{inputenc}
\usepackage{hyperref}
\usepackage{url}
\usepackage{amsmath,amssymb,amsthm}
\usepackage{bm}
\usepackage{booktabs}
\usepackage{enumitem}
\usepackage{graphicx}
\usepackage{microtype}
\usepackage{titletoc}
\usepackage{multirow}
\usepackage{algorithm}
\usepackage{algpseudocode}
\usepackage{float}
\usepackage{xcolor}
\usepackage{listings}
\usepackage{makecell}
\usepackage{adjustbox}  
\usepackage{subcaption}
\definecolor{codeback}{rgb}{0.965,0.965,0.945}
\definecolor{codecomment}{rgb}{0,0.5,0}
\definecolor{codekw}{rgb}{0.55,0,0.55}
\definecolor{codestr}{rgb}{0.25,0.5,0.75}
\definecolor{promptrole}{rgb}{0.10,0.24,0.60}
\definecolor{prompttag}{rgb}{0.80,0.37,0.0}
\lstdefinestyle{promptbox}{
  basicstyle=\ttfamily\scriptsize,
  backgroundcolor=\color{codeback},
  frame=single,
  breaklines=true,
  columns=fullflexible,
  keepspaces=true,
  showstringspaces=false,
  captionpos=b,
  aboveskip=4pt, belowskip=2pt,
  xleftmargin=4pt, xrightmargin=4pt,
  morecomment=[l]{\#},
  commentstyle=\color{codecomment},
  morekeywords={def,return,None,True,False},
  keywordstyle=\color{codekw}\bfseries,
  emph={SYSTEM,USER,MODEL,OUTPUT},
  emphstyle={\color{promptrole}\bfseries},
  emph={[2]TASK,ENV,CONTRACT,METRIC,MENU,SANDBOX,BASIS,DIVERSITY,PREFERENCE,EVENT,DICT,STATE,SCENE,VARIABILITY,FIXED,FITNESS,FROZEN,SKILLS,OBJECTIVE,MAP,ROAD,NETWORK,DEMAND,ORDERS,SIMULATED,TIME,DEPLOYMENT,SUPPLY,FLEET,READ,REQUIREMENT,CANDIDATES},
  emphstyle={[2]\color{prompttag}\bfseries},
}
\lstdefinestyle{pybox}{
  language=Python,
  basicstyle=\ttfamily\scriptsize,
  backgroundcolor=\color{codeback},
  commentstyle=\color{codecomment},
  keywordstyle=\color{codekw}\bfseries,
  stringstyle=\color{codestr},
  frame=single,
  breaklines=true,
  showstringspaces=false,
  captionpos=b,
  aboveskip=4pt, belowskip=2pt,
  xleftmargin=4pt, xrightmargin=4pt,
}

\newcommand{\step}{\mathrm{step}}
\newcommand{\obs}{x}
\newcommand{\phiep}{\phi_{\mathrm{ep}}}
\newcommand{\phistep}{\phi_{\mathrm{step}}}
\newcommand{\ret}{\mathrm{G}}
\newcommand{\task}{\mathcal{M}}

\title{RideSkill: A Hierarchical Algorithm \\ for Generalized Ride Sharing  with \\ LLM-Driven Automatic Evolution}

\author{Zijian Zhao$^{1,2}$, Xialiang Tong$^2$\thanks{Project Lead: Xialiang Tong}, Sen Li$^{1,3}$\thanks{Corresponding Author: Sen Li}, Mingxuan Yuan$^2$  \\
$^1$The Hong Kong University of Science and Technology \\
$^2$Noah's Ark Lab, Huawei \\
$^3$The Hong Kong University of Science and Technology (Guangzhou)
}

\begin{document}
\maketitle

\begin{abstract}
Ride-sharing, which allows multiple passengers with different origin-destination (OD) pairs to share a single vehicle, is a challenging operational problem, as it requires orders with different OD pairs to be efficiently bundled and assigned to vehicles under uncertain and varying scenarios. Although multi-agent reinforcement learning (MARL) solutions have achieved promising performance, they suffer from limited generalization (adapting to different environmental scenarios), low transferability (adapting to different platform objectives), and training difficulties in large-scale systems, such as the curse of dimensionality. Recently, motivated by the scaling of large language models (LLMs), several works have incorporated LLMs into ride-hailing systems, either by employing LLMs directly as decision-making agents or using them for automatic algorithm design. However, none of these approaches support vehicle sharing, which complicates the problem by expanding both the state and action spaces exponentially. Moreover, most of them require frequent LLM calls at inference time, making them infeasible for real-time deployment. To address these issues, we propose RideSkill, a hierarchical method for ride-sharing that leverages LLM-assisted automatic algorithmic design. RideSkill consists of a combiner that assigns appropriate skills to each vehicle from a learned skill repository, enabling adaptive dispatch under varying scenarios and objectives, and a repositioner that sequentially relocates idle vehicles to emerging regions, avoiding conflicts among vehicles. Crucially, the skill repository, combiner, and repositioner are all trained by an LLM-based automatic evolutionary method, eliminating the need for LLM calls during deployment and thus ensuring high real-time performance. Evaluated on a real-world ride-hailing dataset, RideSkill achieves the best performance compared to model-based, MARL-based, and LLM-based benchmarks. Most crucially, our method demonstrates strong generalization and flexibility across different fleet sizes, vehicle speeds, passenger capacities, order densities, and objectives, suggesting high practical application value. Due to company policy, we regret that we cannot release our code.
\end{abstract}

\section{Introduction}

Ride-sharing systems, which allow multiple passengers with different origin–destination (OD) pairs to share a single vehicle, have emerged as a transformative paradigm in urban mobility. By increasing vehicle occupancy and reducing the total number of trips, they alleviate traffic congestion, lower carbon emissions, and improve the overall efficiency of transportation networks. Beyond these environmental benefits, ride-sharing enhances passenger convenience through lower fares and shorter waiting times while enabling platforms to serve more demand with a limited fleet, thereby generating substantial social and economic value in densely populated cities \citep{jin2018ridesourcing}.

Existing approaches to order dispatching in ride-sharing largely fall into two categories. Classical optimization and rule-based methods, although computationally efficient, are typically myopic \citep{alonso2017demand}. Multi-agent reinforcement learning (MARL) techniques have demonstrated stronger performance by learning long-term policies, yet they suffer from the curse of dimensionality when the number of vehicles and orders grows large \citep{hao2022hierarchical}. More critically, once the model is formulated or the policy is trained, they struggle to adapt to changing environmental conditions (e.g., different fleet sizes, vehicle speeds, or capacities) or to shifts in platform objectives. In practice, a platform's objectives (e.g., maximizing service rate, minimizing passenger detour, or balancing vehicle income) may evolve rapidly with market conditions and government regulation, rendering static solutions inadequate. More seriously, most transfer RL methods fail in multi-agent settings, since the change in other agents' policies alters the environment as experienced by any given agent \citep{barreto2018transfer}.

Recently, large language models (LLMs) have been introduced into ride-hailing systems, either by treating an LLM as a decision-making agent that directly selects orders for individual vehicles or by employing it as a global matcher \citep{lyu2026llm,zhang2026large,su2025data}. While these approaches leverage the reasoning capabilities of LLMs, they require frequent model calls at inference time and are therefore unsuitable for the stringent latency requirements of real-time ride-sharing. An alternative line of work uses LLMs for automatic algorithm discovery: by combining the generative power of LLMs with evolutionary search, new heuristics or policies can be designed offline \citep{zhang2026hierarchical}. Compared with traditional evolutionary algorithms, LLMs inject rich prior knowledge and enable more effective exploration of the algorithm space. However, existing LLM-based methods have been developed exclusively for ride-hailing. Extending them to ride-sharing is non-trivial, as it requires considering feasible order combinations, which expands the state and action spaces, modeling inter-order dependencies, and handling vehicles that become heterogeneous once they carry en-route passengers. More critically, all of these methods are still trained on a single fixed scenario and objective, as in conventional MARL-based solutions. As a result, the obtained policies exhibit limited generalization across different operating conditions and limited transferability when the platform's objective changes. (A detailed literature review is provided at Appendix \ref{app:literature}.)

To address these challenges, we propose RideSkill, a hierarchical framework that performs adaptive order dispatching and vehicle repositioning under varying scenarios and platform objectives. RideSkill maintains (i) a repository of reusable basic skills (e.g., nearest matching, detour minimization, service-rate maximization); (ii) a combiner that, conditioned on the current environment and objective, dynamically assigns an appropriate skill to each vehicle; and (iii) a sequential repositioner that relocates idle vehicles to emerging demand regions while avoiding conflicts that would arise from simultaneous decisions. All three components are trained offline by a carefully designed LLM-assisted evolutionary procedure that incorporates self-checking, diverse task generation, and a group-relative fitness formulation. Once trained, the resulting policies operate without any LLM calls, thereby satisfying real-time constraints while retaining strong generalization and transferability. We validate RideSkill on a large-scale ride-sharing simulator built from real New York City trip data. The results show that RideSkill outperforms all types of baselines across different scenarios and objectives. To the best of our knowledge, RideSkill is the first LLM-based solution for ride-sharing tasks, and the first generalized approach capable of addressing diverse scenarios and objectives.

\section{Problem Setup}
\label{sec:setup}

In this paper, we consider a centralized ride-sharing platform operating a fleet of vehicles. Passengers send requests to the platform at arbitrary times, and the platform processes them every $\Delta t$ time units (i.e., the interval between consecutive decision steps) for improved planning. The platform must decide how to dispatch requests (orders) to vehicles, taking into account the OD and temporal relationships between newly assigned orders and en-route orders, the spatial relationships between vehicles and orders, the remaining capacity of vehicles, and the potential impact of future orders. Orders that are not successfully assigned are returned to the order pool for future processing. However, passengers are impatient: if an order is not confirmed within a time threshold, it is canceled, resulting in potential revenue loss and user churn for the platform.

Following \citep{zhao2026ridegym}, we formulate this problem as a Multi-Agent Markov Decision Process (MAMDP), denoted by $\task = \langle n, S, U, \mathcal{P}, \mathrm{R}, \gamma, O, T \rangle$, where $n$ is the number of vehicles (agents), and the remaining components represent the joint state, joint action, state-transition function, reward function, discount factor, joint observation, and time horizon, respectively. We index vehicles by $i \in \mathcal{I} = \{1, \ldots, n\}$ and the orders pending at step $t$ by $j \in \mathcal{J}_t$. Each vehicle $i$ has a local observation $\obs_i$ consisting of its own state (location, status, remaining capacity, committed passengers) and the global request pool. (The detailed notion description is provided at Appendix \ref{app:notation}.)

In conventional solutions, $\task$ is treated as fixed, and policies or models are designed or trained accordingly. In practice, however, $\task$ varies over time. For instance, during peak versus off-peak hours, the platform may operate with different fleet sizes and order volumes, and traffic conditions may affect vehicle speeds. Moreover, as market conditions evolve (e.g., due to inter-platform competition), the objective $\mathrm{R}$ may shift from revenue maximization to service-quality maximization. In this paper, we aim to develop a flexible method that can adapt to these varying tasks (each characterized by a different $\task$) without requiring policy reconstruction or retraining, which would entail substantial computational cost and time.

\section{Methodology}

\subsection{Overview}

\begin{figure}[htbp]
\centering
\includegraphics[width=0.8\textwidth]{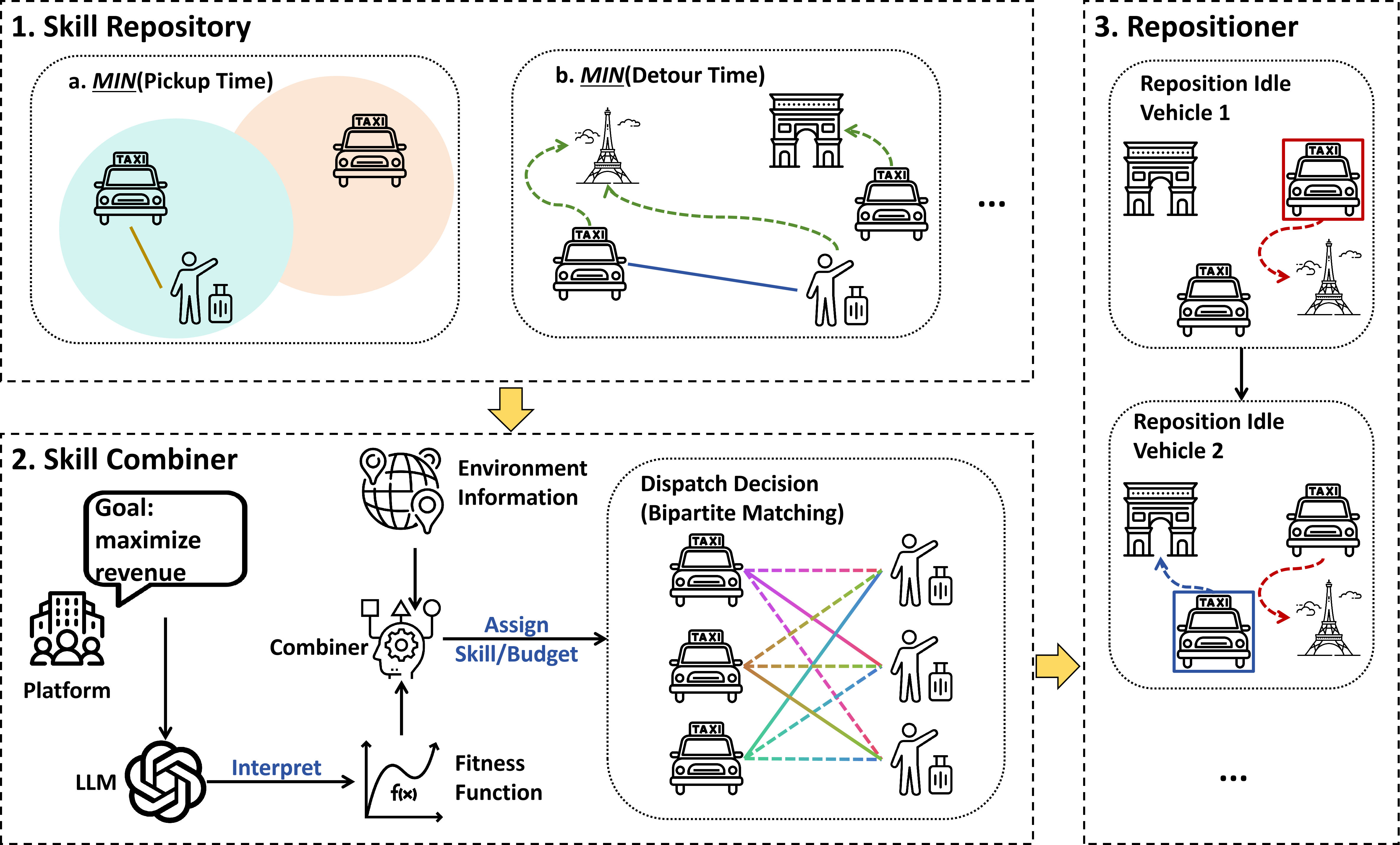}
\caption{Components of RideSkill. A platform objective and environmental context are interpreted by the combiner, which assigns skill weights to each vehicle. A repositioner then sequentially relocates idle vehicles to emerging demand regions.}
\label{fig:main}
\end{figure}

The goal of our method is to design an order-dispatch approach that adapts to varying scenarios and platform objectives. However, end-to-end learning is infeasible, because incorporating such conditions into an already large action and observation space dramatically expands the search space. To address this, we propose a hierarchical solution, illustrated in Figure~\ref{fig:main}. (i) First, a repository of trained basic skills is provided, such as nearest matching, minimizing detour time, and maximizing service rate, which are commonly reusable across different circumstances. (ii) Then, a combiner, conditioned on the current scenario and objective, assigns appropriate skills to each vehicle. For instance, an idle vehicle may be assigned a nearest-matching skill to maximize service rate, while a vehicle with many en-route orders may receive a detour-minimizing skill to ensure service quality. (iii) Finally, a repositioner sequentially processes each idle vehicle and relocates it to regions where it is needed. By operating in a sequential manner, the repositioner implements a step-by-step strategy (analogous to chain-of-thought reasoning), avoiding the risk of relocating too many vehicles to the same region that arises from simultaneous processing.

To optimize these three components, we introduce a four-stage LLM-based evolutionary algorithm for automatic design, shown in Figure~\ref{fig:train}. Built upon the $(\mu+\lambda)$-ES \citep{schwefel1981numerical}, our method leverages LLMs for child generation to exploit their scaling capacity and general knowledge. A self-check module is incorporated, in which the LLM reviews and corrects its outputs to ensure alignment with its own policy and objective design motivations, thereby reducing evolutionary bias. To enhance generalization and transferability, we devise a task-generation mechanism that uses environment randomization alongside LLM-designed objectives to create a diverse training set. Finally, to address the varying reward scales across tasks, we propose a GDPO-style \citep{liu2026gdpo} fitness formulation that incorporates the concept of group advantage from reinforcement learning into the evolutionary algorithm.

In this section, we first introduce the three components and briefly describe how we use the ES algorithm to train them. Then we detail the general LLM-based $(\mu+\lambda)$-ES framework utilized in the training of the three components. The detailed prompt design and algorithm process are provided at Appendix \ref{app:imp}.

\subsection{Components}
\label{sec:components}

\paragraph{Skill Repository}
Recently, the concept of a skill, consisting of a clear textual description, fixed input and output definitions, and a detailed program (e.g., the code or function), has shown high potential for LLMs, with the advantages of being modular, reusable, composable, and standardized \citep{zhou2026comprehensive}. Inspired by this motivation, we maintain a ride-sharing skill repository that can be utilized or combined by drivers (i.e., agents). In the repository, there is a series of basic atomic skills explored and evolved by the LLM, such as nearest matching. The skill repository is denoted by $\mathcal{B} = \{s_1, \ldots, s_K\}$, where each skill $s_k$ has a self-contained order-dispatch policy accompanied by a natural-language card (objective, description, mechanism), so a human can audit it for interpretability.

Due to the large joint action space in the ride-sharing dispatch task, it is difficult to directly output the order assigned to each vehicle. In the current MARL paradigms, we mostly first score each vehicle-order pair (e.g., the Q-value \citep{xu2018large} or the matching probability \citep{zhao2025one}) and then use bipartite matching to maximize the global score. As a result, we believe what does matter is the score function, and we formalize the policy in each skill as a function to score each vehicle-order pair. Formally, a skill $s_k$ maps the observation $\obs_i$ of vehicle $i$, a candidate order $o_j$, the episode-static context $\phiep$ (e.g., road network, fleet size, region layout), and the live per-step context $\phistep$ (e.g., pending counts, and the per-region demand/supply state $\kappa$ introduced below) to a real-valued marginal value for that vehicle-order pair:
\begin{equation}
\label{eq:skill}
    s_k\bigl(\obs_i,\, o_j,\, \phiep,\, \phistep\bigr) \in \mathbb{R}.
\end{equation}
The per-pair scores produced by the skills are converted into a concrete dispatch by solving a bipartite matching problem, formulated as an integer linear program that assigns each order to at most one vehicle and each vehicle to at most one new order per decision step:
\begin{subequations}
\label{eq:matching}
\begin{align}
& \max_{\{u_{i,j}\}} \sum_{i \in \mathcal{I}} \sum_{j \in \mathcal{J}_t \cup \{\varnothing\}} u_{i,j} \cdot a_{i,j}, \\
\text{s.t.}\quad
& \sum_{i \in \mathcal{I}} u_{i,j} \le 1, \quad \forall\, j \in \mathcal{J}_t, \\
& \sum_{j \in \mathcal{J}_t \cup \{\varnothing\}} u_{i,j} \ge 1, \quad \forall\, i \in \mathcal{I}, \\
& u_{i,\varnothing} \cdot \sum_{j \in \mathcal{J}_t} u_{i,j} = 0, \quad \forall\, i \in \mathcal{I}, \\
& \sum_{j \in \mathcal{J}_t} u_{i,j} \cdot h_{j} \le c_{i}, \quad \forall\, i \in \mathcal{I}, \\
& u_{i,j} \in \{0, 1\}, \quad \forall\, i \in \mathcal{I},\ \forall\, j \in \mathcal{J}_t \cup \{\varnothing\},
\end{align}
\end{subequations}
where $u_{i,j}$ is the assignment variable of vehicle $i$ to order $j$, $a_{i,j}$ is the matching score of pair $(i,j)$ (a single skill's score in Eq.~\eqref{eq:skill}, or the blended score in Eq.~\eqref{eq:blended} below; set to $-\infty$ for infeasible pairs), $\varnothing$ is a dummy order representing the no-op option that enables a vehicle to wait, $h_j$ is the passenger count of order $j$, and $c_i$ is the remaining capacity of vehicle $i$ at the current step.

During offline training, each skill is authored and trained entirely by the LLM. Specifically, the LLM is asked to explore various skills by itself, where each skill should have a different objective and not be redundant to existing ones. In each iteration, the LLM proposes a new skill together with its objective, mechanism, and a fitness function over rollout metrics. The proposal is compiled, sandbox-validated, and evolved by our LLM-based $(\mu{+}\lambda)$-ES algorithm. Once a skill is well trained, its evaluation metric is compared with the existing ones (through a similarity metric such as cosine similarity) to prevent near-duplicates from accumulating. The loop stops when the repository reaches $K$ skills or when $N_f$ consecutive rounds fail to produce a new distinct behavior.

\paragraph{Combiner}
As in a multi-agent LLM system, different vehicles have different situations and require suitable skills accordingly. To realize this, we introduce a combiner that assigns appropriate skills to agents conditioned on their state, the environment condition, and the objective. Formally, the combiner $\pi_c$ maps the current contexts and the objective to a real-valued score for each frozen skill:
\begin{equation}
\label{eq:combiner}
    \pi_c\bigl(\obs_i,\, \phiep,\, \phistep,\, w\bigr) = \alpha_i = (\alpha_{i,1}, \ldots, \alpha_{i,K}) \in \mathbb{R}^{K},
\end{equation}
where $\alpha_{i,k} \in \mathbb{R}$ is the raw score of skill $k$ for vehicle $i$. These scores are converted into blend weights at dispatch time by keeping the $b$ highest-positive skills, $\mathcal{K}_i = \mathrm{top}\text{-}b\!\bigl(\{k : \alpha_{i,k} > 0\}\bigr)$, and applying a softmax:
\begin{equation}
\label{eq:blend-weights}
    \tilde{\alpha}_{i,k} = \frac{\exp(\alpha_{i,k})}{\sum_{k' \in \mathcal{K}_i} \exp(\alpha_{i,k'})}, \quad k \in \mathcal{K}_i,
\end{equation}
so that $\tilde\alpha_i = (\tilde{\alpha}_{i,k})_{k \in \mathcal{K}_i}$ lies on the simplex over $\mathcal{K}_i$. Here $w$ denotes the platform objective, encoded as a per-step reward function that maps a per-vehicle event descriptor (e.g., the orders assigned or completed at this step, together with their waiting, service, and detour times) to a scalar reward. Since the objective is provided by the user, it may take unexpected forms or even be a black box, so it is difficult to encode such a function into the combiner directly. To address this, we propose an event-detection mechanism, in which the LLM evolves a set of basic probe events (e.g., assigning an order to a vehicle with en-route orders) alongside the combiner; the combiner calls $w$ on these events and reads the differences between paired responses to estimate the shape of the reward function. During task generation, the coefficients of each authored objective are normalized to sum to one (a single global rescale that preserves the term ratios; see Section~\ref{sec:training}), so the probed responses are comparable across objectives with different reward scales.

The resulting blend weights are used to score each pending order, normalized per skill over the vehicle's candidate set:
\begin{equation}
\label{eq:blended}
    a_{i,j} = \sum_{k \in \mathcal{K}_i} \tilde{\alpha}_{i,k} \cdot \frac{s_k(\obs_i, o_j, \phiep, \phistep) - \mu_k}{\sigma_k + \epsilon},
\end{equation}
where $\mu_k$ and $\sigma_k$ are the mean and standard deviation of skill $k$'s scores over vehicle $i$'s feasible candidate set $\mathcal{J}_i \subseteq \mathcal{J}_t$ (we omit their dependence on $i$ for brevity), with $\epsilon > 0$ a small constant guarding the degenerate case $\sigma_k = 0$. This per-skill standardization is necessary because different skills produce scores on different ranges, and mixing them directly would let large-range skills annihilate the signals of the others. The combiner is trained by evolving the probing and routing logic over a distribution of objectives and scenarios through the same LLM-based $(\mu{+}\lambda)$-ES, under the well-trained skill repository.

Moreover, inspired by the bid mechanism where each participant has a different budget, we optionally rescale the matching scores to promote fairness among vehicles. The intuition is that to make drivers' rewards (a reflection of their income) more similar, we should give more budget to those with a lower historical cumulative reward. Formally, we compute an income-based budget per vehicle:
\begin{equation}
\label{eq:budget}
    \beta_i = \exp\!\bigl(-\rho \cdot z_i\bigr), \quad z_i = \frac{E_i - \bar{E}}{\sigma_E + \epsilon},
\end{equation}
where $E_i$ is vehicle $i$'s cumulative take-home income, $\bar{E}$ and $\sigma_E$ are the fleet-wide mean and standard deviation of cumulative income, $\rho \ge 0$ is a fairness-strength parameter, and $\epsilon > 0$ is a small constant. Below-average vehicles ($z_i < 0$) receive $\beta_i > 1$, which boosts their matching scores $a_{i,j}$ multiplicatively in Eq.~\eqref{eq:matching}, while above-average vehicles are damped; $\rho = 0$ disables the mechanism entirely. 

\emph{Note that this rescaling mechanism is heuristic and does not come with theoretical guarantees. Empirically, we observe that it is effective without substantially affecting overall performance. Moreover, the mechanism is optional and can be enabled or disabled at the user's discretion.}

\paragraph{Repositioner}
\label{sec:repositioner}
Demand and idle-vehicle supply may be unevenly distributed across space and time, so after the combiner dispatches orders, many vehicles end up idle in low-demand regions, wasting capacity that could serve near-future demand. As a result, we introduce a repositioner that proactively relocates idle vehicles toward regions where pickups are most likely, improving coverage and reducing passenger waiting times before orders even arrive. However, if the repositioner processes all vehicles simultaneously, the large joint state and action space make it difficult to explore a good strategy and may cause too many vehicles to emerge in a single region. To tackle this, we process idle vehicles one at a time in a randomized order: after each vehicle is assigned a target region, that region's effective demand is decremented before the next vehicle decides, which naturally spreads vehicles across regions and prevents the simultaneous flocking that would over-supply a single hotspot. Formally, let $\mathcal{G}$ denote the set of regions and $\kappa = (\kappa_g)_{g \in \mathcal{G}}$ the shared per-region effective demand/supply state, which is a component of $\phistep$ but is passed to the repositioner explicitly because it is mutated during the sequential pass. The repositioner additionally takes the vehicle's fairness budget $\beta_i$, so that it can bias relocation toward boosted vehicles. For each idle vehicle $i$, the repositioner scores every region in a candidate set $\mathcal{G}_i \subseteq \mathcal{G}$ (the vehicle's current region, its neighbors, and the top-$H$ globally hottest regions by live demand) using a learned function:
\begin{equation}
\label{eq:repositioner}
    \pi_r\bigl(\obs_i,\, \phiep,\, \phistep,\, \kappa,\, w,\, \beta_i\bigr) = \{\nu_g\}_{g \in \mathcal{G}_i},
\end{equation}
where $\nu_g \in \mathbb{R}$ is the relocation score of region $g$. The vehicle is sent to the highest-scoring region $g^* = \arg\max_{g \in \mathcal{G}_i} \nu_g$, after which $\kappa_{g^*}$ is decremented. The repositioner is evolved like the combiner, through the same four-stage procedure. Unlike the previous two phases, its fitness is designed as the improvement of the episode return over the reposition-off baseline on the same task and random seed, $\ret^{\mathrm{repo}} - \ret^{\mathrm{off}}$ (Section~\ref{sec:training}), rather than the raw return. Consequently, a positive delta directly indicates that repositioning genuinely improves performance relative to doing nothing.

\subsection{Training Process}
\label{sec:training}

\begin{figure}[htbp]
\centering
\includegraphics[width=0.8\textwidth]{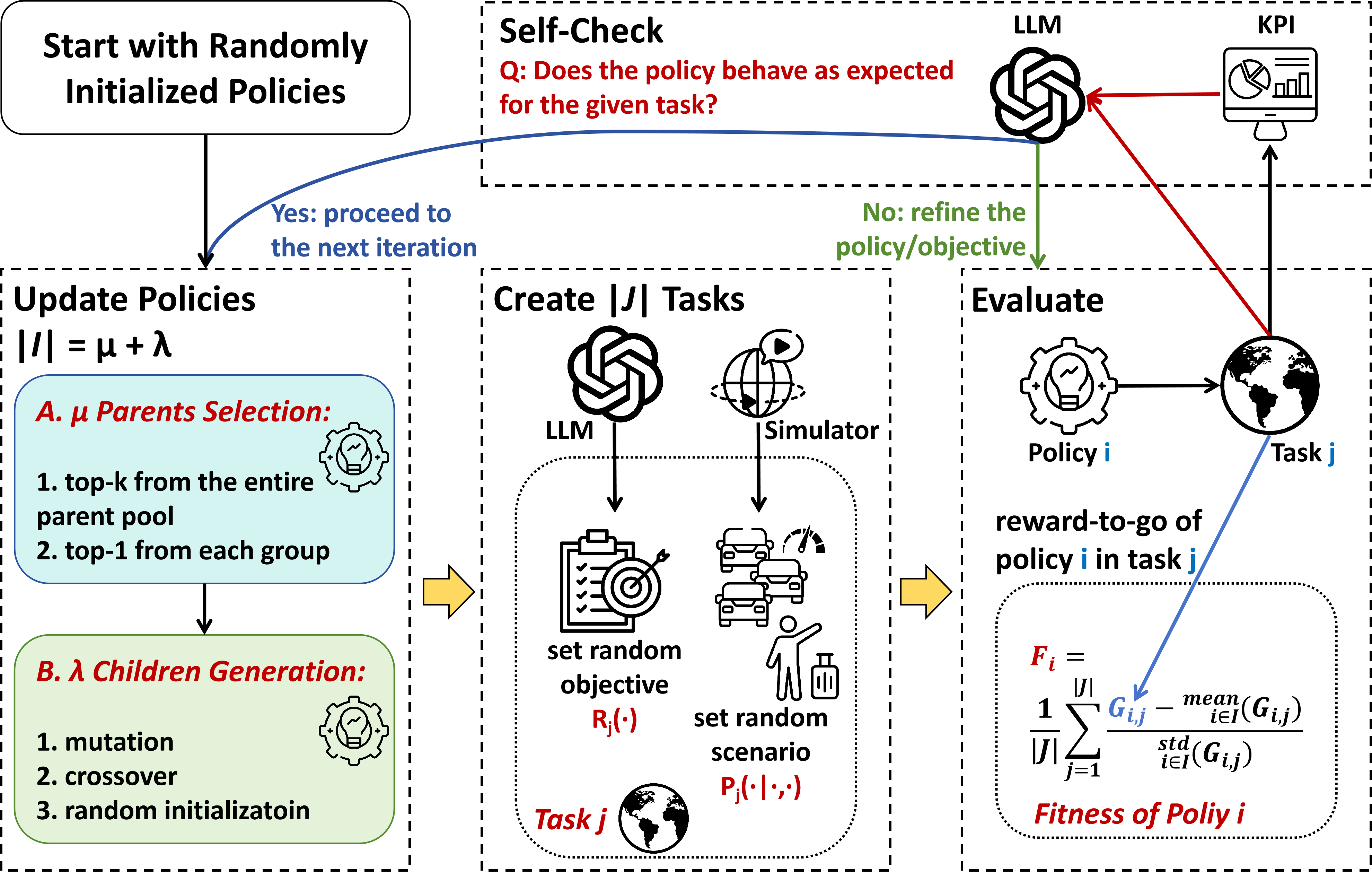}
\caption{LLM-assisted evolutionary training pipeline. Each generation (i) selects $\mu$ parents and generates $\lambda$ children, (ii) evaluates each candidate on a batch of tasks, and computes the group-relative advantage as fitness, and (iii) feeds the result through a self-check before proceeding to the next iteration.}
\label{fig:train}
\end{figure}

To train the three components above, we use the $(\mu{+}\lambda)$-ES~\citep{schwefel1981numerical}. Recently, LLMs have shown high potential in automatic algorithm design \citep{zhang2026hierarchical,su2025data}, where their high understanding, analysis, and general knowledge can help evolve algorithm exploration more efficiently, instead of randomly generating children each iteration, as in conventional ES solutions. The whole process of the LLM-based $(\mu{+}\lambda)$-ES is shown in Figure~\ref{fig:train}, where we made some careful adaptations for our task.

At each generation, the procedure follows three stages. (i) First, $\mu$ parents are selected from the current generation: the top-$\mu$ by fitness, plus, in Phases~2 and~3, one reserved elite per task group (e.g., a family of objectives, or a fairness-strength band, defined by the LLM itself), so that specialists on hard tasks survive as crossover material. (ii) Second, $\lambda$ children are generated from these parents: an LLM crossover of two parents with probability $p_{\mathrm{cross}}$, otherwise a single-parent mutation, plus one parentless fresh injection per generation to encourage variety. Different from conventional ES, these operations are based on the result analysis of the parents instead of random exploration. Every candidate undergoes sandbox compilation and field validation; failures are fed back with progressively lower temperature. (iii) Third, each candidate policy $\pi_l$ is evaluated by rollout on a batch of tasks $\mathcal{T} = \{\task_m\}_{m=1}^{|\mathcal{T}|}$, where each task $\task_m$ comprises a platform reward function $\mathrm{R}_m$ and an environment dynamics model $\mathcal{P}_m$. Each search concludes with a self-check that reviews the measured rollout against the original design intent. The core design of these stages is detailed below.

\paragraph{LLM as the evolutionary operator.} Standard $(\mu{+}\lambda)$-ES uses domain-specific mutation and crossover operators (e.g., Gaussian perturbation of real-valued vectors). We replace these with the LLM, since the LLM's prior knowledge of ride-pooling dispatch, the codebase conventions, and the skill/combiner contracts enables it to produce structurally diverse offspring that go beyond coefficient perturbation. Moreover, we follow the prompt design in \citep{su2025data} to let the LLM generate children targetedly, according to the performance of the parents. This also enables more varied exploration, since the LLM can directly edit the function, whereas conventional methods must rely on some pre-defined operators.

\paragraph{Self-proposed diverse tasks.} In conventional ES and ride-sharing methods, the tasks are pre-fixed by the designer, making the trained algorithm difficult to adapt to out-of-distribution tasks. In our method, the LLM proposes the various objectives and scenarios it trains on by itself. In Phase~1 (skill repository training), the LLM self-invents each skill's objective and writes its own fitness function, so the task batch consists of environment scenarios only. In Phase~2 (combiner training) and Phase~3 (repositioner training), reward functions are drawn from some LLM-authored structurally distinct families, including objectives the model authors from a natural-language brief. Specifically, for each objective, the scenario setup is generated randomly, to push the trained algorithm to adapt to different tasks and to have the capacity to transfer to unseen tasks.

\paragraph{GDPO-style fitness across varying reward scales.}
Since different tasks have vastly different reward magnitudes, we adopt a formulation inspired by GRPO~\citep{liu2026gdpo} that makes the fitness task-invariant. This group-relative fitness is used in Phases~2 and~3, where every candidate is scored by the return of its rollout on a task; Phase~1 instead grades its candidates by the self-authored fitness $f_{\mathrm{self}}$ of the skill under evolution (Section~\ref{sec:components}), which is fixed at generation 0. Let $\ret_l(\task_m)$ denote the return (cumulative reward) obtained by rolling out candidate $\pi_l$ on task $\task_m$. For each candidate $\pi_l$ on task $\task_m$, the group-relative advantage is
\begin{equation}
\label{eq:gra}
    A_{l,m} = \frac{\ret_l(\task_m) - \mu^{(m)}}{\sigma^{(m)}},
\end{equation}
where $\mu^{(m)}$ and $\sigma^{(m)}$ are the mean and standard deviation of the returns $\{\ret_l(\task_m)\}_{l}$ over the entire group of candidates evaluated on task $\task_m$. In Phase~3, the return $\ret_l(\task_m)$ in Eq.~\eqref{eq:gra} is replaced by the paired improvement $\ret_l^{\mathrm{repo}}(\task_m) - \ret^{\mathrm{off}}(\task_m)$, where the two returns are rolled on the same seed with repositioning on (using candidate $\pi_l$) versus off, so what is normalized is precisely how much repositioning helps beyond doing nothing. The fitness of policy $\pi_l$ is then the mean advantage over all tasks in the batch:
\begin{equation}
\label{eq:fitness}
    F_l = \frac{1}{|\mathcal{T}|} \sum_{m=1}^{|\mathcal{T}|} A_{l,m},
\end{equation}
and each search ends with a runoff in which the distinct round-leaders are re-rolled together on a fresh batch to select the final champion (Algorithms~\ref{alg:phase2} and~\ref{alg:phase3}). In this way, we can distinguish the relative performance of policies within each task. On the contrary, if we simply use the sum of the raw returns, the signal of low-reward-range tasks would be annihilated by the large-range ones.

\paragraph{Self-check with feedback loop.}
\label{sec:self-check}
A core idea behind using the LLM for algorithm design and evolution is the high interpretability of each design and process. However, due to bias in knowledge or hallucination, some proposed policy or objective may exhibit different performance from the LLM's intuitive design motivation, making the description untrustworthy. To tackle this problem, we propose a self-check feedback loop applied when a search concludes. It takes two forms depending on the phase. In Phase~1, the skill audit reviews the measured rollout of the champion against the original design intent and returns one of three verdicts: match (freeze as-is), description\_wrong (the skill does something coherent but not what the card claims; rewrite the card), or fitness\_wrong (the authored fitness rewards the wrong behavior; rewrite the fitness and re-search). If the audit fails, the process routes back to child generation for refinement. In Phases~2 and~3, the fitness is fixed, so the failure this audit guards against is instead a program that ignores the objective $w$ entirely; the policy audit measures this as a counterfactual (rolling the same seed with and without handing the program $w$) and records an objective-responsiveness verdict next to the frozen artifact, without feeding back into the search.

\section{Experiments}
\label{sec:experiments}


\paragraph{Simulator and data.} To validate our proposed method, we conduct a series of experiments using real-world ride-hailing data in Manhattan, New York City \citep{TLCData}. The simulation is run on RideGym~\citep{zhao2026ridegym}, a ride-sharing simulator with a standardized Gym API. We use the 8:00 to 20:00 period between April 6 and April 12, 2026 as the training set, and the 8:00 to 20:00 periods on April 13 and April 14 as the validation and testing sets.

\paragraph{Baselines.} We compare against methods of different types: \textbf{(i) model-based methods:} nearest matching \citep{kalyanasundaram1993online}, Kuhn-Munkres (KM) \citep{kuhn1955hungarian,simonetto2019real}, and Gale-Shapley (GS) \citep{gale1962college,yue2024end}; \textbf{(ii) MARL-based methods:} REDA \citep{holder2025multi}, BMG-Q \citep{hu2025bmg}, and MF-DDQN \citep{li2019efficient} (referred to as MFRL in the tables); and \textbf{(iii) LLM-based methods:} \cite{zhang2026hierarchical} (including the open-loop and close-loop versions). For fairness, we use the same Claude Opus 4.8 model \citep{anthropic2026opus48} in both our method and theirs. Here, we do not use the other methods mentioned in Table~\ref{tab:comparison}, since they are difficult to scale to real-world large-scale scenarios. All agents were trained on the anchor objective at fleet=1{,}000, capacity=4, speed=35 km/h.

\paragraph{RideSkill ablations.} To evaluate the effect of each module we designed, we compare the performance of (i) full RideSkill, (ii) RideSkill (w/o reps) in which we eliminate the repositioner so that the comparison to model-based and MARL-based methods is more direct, since they do not have a reposition design, and (iii) single-skill, in which we evaluate the average performance of a single skill designed in the well-trained skill repository, to gauge the effect of our proposed combiner. During deployment, the matcher scores the top-60 candidate orders per vehicle (the feasible candidate set $\mathcal{J}_i$) and blends the top-$b{=}3$ skills.

\paragraph{Experiment Results.}
As shown in Table~\ref{tab:generalization}, we compare the performance of our method against baselines under various scenarios, reporting average results across multiple hours, fleet sizes, speeds, and capacities. For each configuration, only one setting is varied relative to the training conditions used by the benchmarks (where across-hour testing corresponds to an in-domain setting, and the remaining variations are out-of-domain). Our method consistently achieves the best performance across all scenarios, and does so nearly across all metrics. Notably, the detour time is substantially lower than that of competing approaches, indicating that the LLM-based design is effective in finding efficient order-bundling solutions. We also observe that all MARL-based benchmarks fail to handle scenarios where vehicle capacity exceeds the training range, due to their fixed network input sizes. In contrast, our method significantly outperforms both rule-based and LLM-based baselines in these cases. Finally, RideSkill outperforms RideSkill (w/o repo), confirming the effectiveness of our repositioning mechanism. Both variants substantially outperform the single-skill baseline, which suggests that no single strategy can adapt to all scenarios, validating the necessity of our combiner design. This flexibility to different scenarios is a key distinction between our method and prior approaches.  Due to page limitations, we defer detailed analyses of the fairness mechanism, objective adaptation, performance with other LLMs, and trained examples to Appendix~\ref{app:exp}.

\begin{table}[t!]
\centering
\caption{Mean$\pm$std across all levels of the given axis. Hour: 08--19 (12 levels); fleet: 200--1{,}500 (7 levels); speed: 20--50 km/h (7 levels); capacity $\le$ 4: 1--4 (4 levels); capacity $>$ 4: 5--8 (3 levels, MARL excluded). Anchor objective; other variables held at in-domain. Bold is best and underline is second-best per column, a convention that applies to all subsequent tables.}
\label{tab:generalization}
\smallskip
\adjustbox{max width=\textwidth}{%
\begin{tabular}{@{} l l r r r r r r r @{}}
\toprule
& \textbf{Method} & \textbf{Reward} & \textbf{Service} & \textbf{Complete} & \textbf{Wait} & \textbf{Ride} & \textbf{Detour} & \textbf{Util} \\
\midrule
\multirow{11}{*}{\rotatebox{90}{\small\textbf{Hour}}}
& nearest & 6{,}051$\pm$906 & 0.90$\pm$0.06 & 0.75$\pm$0.06 & 1.72$\pm$0.37 & 7.00$\pm$0.24 & 1.77$\pm$0.30 & 0.72$\pm$0.14 \\
& KM & 5{,}765$\pm$764 & 0.88$\pm$0.08 & 0.74$\pm$0.08 & 1.59$\pm$0.27 & 7.08$\pm$0.27 & 1.90$\pm$0.34 & 0.68$\pm$0.11 \\
& GS & 5{,}480$\pm$693 & 0.85$\pm$0.12 & 0.70$\pm$0.12 & 1.64$\pm$0.28 & 7.35$\pm$0.58 & 1.92$\pm$0.36 & 0.67$\pm$0.11 \\
& REDA & 6{,}743$\pm$998 & \underline{0.92}$\pm$\underline{0.06} & 0.79$\pm$0.08 & 1.44$\pm$0.13 & 6.60$\pm$0.68 & 1.15$\pm$0.64 & 0.78$\pm$0.14 \\
& MFRL & 6{,}850$\pm$1{,}042 & \textbf{0.93}$\pm$\textbf{0.06} & \underline{0.81}$\pm$\underline{0.08} & 1.51$\pm$0.10 & 6.53$\pm$0.65 & 1.13$\pm$0.61 & 0.80$\pm$0.14 \\
& BMG-Q & 6{,}657$\pm$1{,}008 & \underline{0.92}$\pm$\underline{0.07} & 0.79$\pm$0.08 & 1.46$\pm$0.06 & 6.79$\pm$0.53 & 1.38$\pm$0.47 & 0.78$\pm$0.14 \\
& \cite{zhang2026hierarchical} (open) & 5{,}780$\pm$787 & 0.88$\pm$0.07 & 0.73$\pm$0.07 & 1.86$\pm$0.46 & 7.12$\pm$0.25 & 1.90$\pm$0.31 & 0.70$\pm$0.14 \\
& \cite{zhang2026hierarchical} (close) & 5{,}835$\pm$811 & 0.88$\pm$0.07 & 0.74$\pm$0.07 & 1.83$\pm$0.45 & 7.06$\pm$0.22 & 1.84$\pm$0.28 & 0.70$\pm$0.14 \\
& single-skill & 4{,}876$\pm$970 & 0.72$\pm$0.05 & 0.61$\pm$0.04 & 2.35$\pm$0.08 & 5.91$\pm$0.20 & 1.12$\pm$0.13 & 0.61$\pm$0.11 \\
& RideSkill (w/o reps) & \underline{7{,}221}$\pm$\underline{1{,}360} & 0.88$\pm$0.07 & 0.79$\pm$0.06 & \underline{1.14}$\pm$\underline{0.07} & \textbf{5.05}$\pm$\textbf{0.27} & \underline{0.17}$\pm$\underline{0.02} & \underline{0.80}$\pm$\underline{0.10} \\
& {RideSkill} & \textbf{7{,}654}$\pm$\textbf{1{,}497} & \underline{0.92}$\pm$\underline{0.06} & \textbf{0.83}$\pm$\textbf{0.05} & \textbf{1.12}$\pm$\textbf{0.15} & \underline{5.24}$\pm$\underline{0.24} & \textbf{0.16}$\pm$\textbf{0.02} & \textbf{0.90}$\pm$\textbf{0.10} \\
\midrule
\multirow{11}{*}{\rotatebox{90}{\small\textbf{Fleet}}}
& nearest & 5{,}288$\pm$2{,}801 & 0.63$\pm$0.24 & 0.52$\pm$0.21 & 2.25$\pm$0.30 & 7.63$\pm$0.47 & 2.66$\pm$0.62 & 0.90$\pm$0.08 \\
& KM & 4{,}981$\pm$2{,}467 & 0.61$\pm$0.23 & 0.50$\pm$0.20 & 1.98$\pm$0.22 & 7.64$\pm$0.35 & 2.69$\pm$0.46 & 0.84$\pm$0.10 \\
& GS & 4{,}390$\pm$2{,}405 & 0.50$\pm$0.25 & 0.40$\pm$0.21 & 2.42$\pm$0.71 & 9.38$\pm$1.80 & 2.60$\pm$0.39 & 0.82$\pm$0.10 \\
& REDA & 5{,}681$\pm$3{,}245 & 0.64$\pm$0.25 & 0.53$\pm$0.23 & 1.73$\pm$0.29 & 8.06$\pm$0.91 & 2.76$\pm$1.02 & {0.92}$\pm${0.06} \\
& MFRL & 5{,}761$\pm$3{,}377 & \underline{0.65}$\pm$\underline{0.26} & {0.54}$\pm${0.24} & 1.76$\pm$0.27 & 8.00$\pm$0.96 & 2.73$\pm$1.06 & \underline{0.94}$\pm$\underline{0.04} \\
& BMG-Q & 5{,}676$\pm$3{,}200 & 0.64$\pm$0.25 & 0.53$\pm$0.22 & {1.62}$\pm${0.23} & 8.01$\pm$0.84 & 2.77$\pm$1.00 & 0.92$\pm$0.05 \\
& \cite{zhang2026hierarchical} (open) & 4{,}842$\pm$2{,}753 & 0.59$\pm$0.26 & 0.48$\pm$0.22 & 3.14$\pm$1.07 & 7.60$\pm$0.33 & 2.52$\pm$0.34 & 0.88$\pm$0.09 \\
& \cite{zhang2026hierarchical} (close) & 4{,}963$\pm$2{,}729 & 0.60$\pm$0.25 & 0.49$\pm$0.22 & 2.91$\pm$0.81 & 7.40$\pm$0.18 & 2.44$\pm$0.29 & 0.88$\pm$0.09 \\
& single-skill & 4{,}835$\pm$1{,}862 & 0.49$\pm$0.18 & 0.42$\pm$0.16 & 2.06$\pm$0.06 & 6.16$\pm$0.25 & 1.36$\pm$0.16 & 0.73$\pm$0.07 \\
& RideSkill (w/o reps) & \underline{7{,}517}$\pm$\underline{2{,}638} & 0.63$\pm$0.21 & \underline{0.58}$\pm$\underline{0.19} & \textbf{1.16}$\pm$\textbf{0.08} & \textbf{4.14}$\pm$\textbf{0.68} & \textbf{0.19}$\pm$\textbf{0.01} & 0.85$\pm$0.08 \\
& {RideSkill} & \textbf{7{,}907}$\pm$\textbf{2{,}945} & \textbf{0.66}$\pm$\textbf{0.24} & \textbf{0.61}$\pm$\textbf{0.21} & \underline{1.22}$\pm$\underline{0.10} & \underline{4.24}$\pm$\underline{0.80} & \textbf{0.19}$\pm$\textbf{0.02} & \textbf{0.95}$\pm$\textbf{0.02} \\
\midrule
\multirow{11}{*}{\rotatebox{90}{\small\textbf{Speed}}}
& nearest & 6{,}701$\pm$2{,}158 & 0.77$\pm$0.14 & 0.63$\pm$0.16 & 2.41$\pm$0.85 & 7.72$\pm$1.96 & 2.30$\pm$0.50 & 0.89$\pm$0.02 \\
& KM & 6{,}135$\pm$1{,}710 & 0.73$\pm$0.12 & 0.59$\pm$0.14 & 1.97$\pm$0.50 & 7.79$\pm$1.83 & 2.41$\pm$0.35 & 0.79$\pm$0.02 \\
& GS & 5{,}429$\pm$1{,}557 & 0.63$\pm$0.13 & 0.50$\pm$0.14 & 2.11$\pm$0.58 & 8.84$\pm$2.34 & 2.44$\pm$0.30 & 0.77$\pm$0.02 \\
& REDA & 7{,}078$\pm$2{,}349 & 0.78$\pm$0.13 & 0.64$\pm$0.16 & 1.75$\pm$0.67 & 8.02$\pm$2.33 & 2.26$\pm$0.76 & 0.91$\pm$0.02 \\
& MFRL & 7{,}165$\pm$2{,}466 & \textbf{0.79}$\pm$\textbf{0.14} & {0.65}$\pm${0.17} & 1.84$\pm$0.73 & 7.96$\pm$2.35 & 2.23$\pm$0.79 & \underline{0.93}$\pm$\underline{0.01} \\
& BMG-Q & 7{,}098$\pm$2{,}280 & 0.78$\pm$0.13 & 0.64$\pm$0.16 & {1.65}$\pm${0.58} & 8.02$\pm$2.30 & 2.29$\pm$0.75 & 0.91$\pm$0.02 \\
& \cite{zhang2026hierarchical} (open) & 6{,}272$\pm$2{,}106 & 0.74$\pm$0.14 & 0.60$\pm$0.17 & 2.81$\pm$1.03 & 7.80$\pm$1.87 & 2.35$\pm$0.39 & 0.87$\pm$0.03 \\
& \cite{zhang2026hierarchical} (close) & 6{,}369$\pm$2{,}061 & 0.75$\pm$0.14 & 0.61$\pm$0.16 & 2.74$\pm$0.96 & 7.69$\pm$1.81 & 2.28$\pm$0.36 & 0.87$\pm$0.03 \\
& single-skill & 5{,}780$\pm$1{,}126 & 0.59$\pm$0.09 & 0.50$\pm$0.11 & 2.24$\pm$0.71 & 6.38$\pm$1.68 & 1.27$\pm$0.18 & 0.70$\pm$0.03 \\
& RideSkill (w/o reps) & \underline{8{,}904}$\pm$\underline{1{,}151} & 0.75$\pm$0.11 & \underline{0.68}$\pm$\underline{0.12} & \textbf{1.19}$\pm$\textbf{0.31} & \textbf{4.81}$\pm$\textbf{1.13} & \textbf{0.20}$\pm$\textbf{0.04} & {0.82}$\pm${0.04} \\
& {RideSkill} & \textbf{9{,}457}$\pm$\textbf{1{,}371} & \textbf{0.79}$\pm$\textbf{0.13} & \textbf{0.72}$\pm$\textbf{0.13} & \underline{1.32}$\pm$\underline{0.38} & \underline{4.96}$\pm$\underline{1.07} & \textbf{0.20}$\pm$\textbf{0.05} & \textbf{0.95}$\pm$\textbf{0.00} \\
\midrule
\multirow{11}{*}{\rotatebox{90}{\small\textbf{Cap$\leq$4}}}
& nearest & 5{,}193$\pm$1{,}641 & 0.54$\pm$0.21 & 0.46$\pm$0.16 & 1.70$\pm$0.42 & 6.55$\pm$0.72 & 1.34$\pm$0.79 & 0.65$\pm$0.23 \\
& KM & 4{,}877$\pm$1{,}368 & 0.52$\pm$0.19 & 0.44$\pm$0.15 & 1.55$\pm$0.30 & 6.60$\pm$0.75 & 1.44$\pm$0.86 & 0.60$\pm$0.19 \\
& GS & 4{,}527$\pm$1{,}107 & 0.48$\pm$0.15 & 0.40$\pm$0.11 & 1.61$\pm$0.32 & 6.96$\pm$1.08 & 1.46$\pm$0.86 & 0.59$\pm$0.18 \\
& REDA & 5{,}548$\pm$1{,}845 & 0.55$\pm$0.21 & 0.47$\pm$0.17 & {1.37}$\pm${0.17} & 6.53$\pm$0.85 & 1.08$\pm$0.85 & 0.68$\pm$0.24 \\
& MFRL & {5{,}714}$\pm${1{,}912} & \textbf{0.56}$\pm$\textbf{0.21} & \underline{0.49}$\pm$\underline{0.17} & 1.55$\pm$0.05 & {6.41}$\pm${0.86} & {1.04}$\pm${0.83} & \underline{0.71}$\pm$\underline{0.24} \\
& BMG-Q & 5{,}580$\pm$1{,}835 & 0.55$\pm$0.21 & 0.47$\pm$0.16 & 1.51$\pm$0.03 & 6.50$\pm$0.85 & 1.11$\pm$0.81 & {0.69}$\pm${0.23} \\
& \cite{zhang2026hierarchical} (open) & 4{,}946$\pm$1{,}476 & 0.53$\pm$0.20 & 0.45$\pm$0.15 & 1.89$\pm$0.52 & 6.64$\pm$0.72 & 1.43$\pm$0.81 & 0.63$\pm$0.22 \\
& \cite{zhang2026hierarchical} (close) & 5{,}010$\pm$1{,}499 & 0.53$\pm$0.20 & 0.45$\pm$0.15 & 1.85$\pm$0.50 & 6.59$\pm$0.68 & 1.39$\pm$0.79 & 0.64$\pm$0.22 \\
& single-skill & 4{,}087$\pm$1{,}531 & 0.41$\pm$0.16 & 0.36$\pm$0.13 & 1.79$\pm$0.18 & 5.60$\pm$0.36 & 0.81$\pm$0.43 & 0.51$\pm$0.18 \\
& RideSkill (w/o reps) & \underline{6{,}142}$\pm$\underline{2{,}422} & 0.53$\pm$0.20 & 0.48$\pm$0.18 & \textbf{1.16}$\pm$\textbf{0.06} & \textbf{5.00}$\pm$\textbf{0.32} & \underline{0.18}$\pm$\underline{0.02} & 0.61$\pm$0.19 \\
& {RideSkill} & \textbf{6{,}538}$\pm$\textbf{2{,}667} & \textbf{0.56}$\pm$\textbf{0.22} & \textbf{0.51}$\pm$\textbf{0.20} & \underline{1.22}$\pm$\underline{0.11} & \underline{5.17}$\pm$\underline{0.27} & \textbf{0.17}$\pm$\textbf{0.02} & \textbf{0.93}$\pm$\textbf{0.02} \\
\midrule
\multirow{8}{*}{\rotatebox{90}{\small\textbf{Cap$>$4}}}
& nearest & 5{,}970$\pm$617 & \textbf{0.86}$\pm$\textbf{0.03} & 0.67$\pm$0.00 & 2.02$\pm$0.15 & 8.34$\pm$0.45 & 3.46$\pm$0.55 & \underline{0.83}$\pm$\underline{0.05} \\
& KM & 5{,}333$\pm$532 & {0.83}$\pm${0.04} & 0.64$\pm$0.01 & 1.83$\pm$0.04 & 8.44$\pm$0.46 & 3.62$\pm$0.54 & 0.76$\pm$0.02 \\
& GS & 5{,}038$\pm$352 & 0.77$\pm$0.07 & 0.59$\pm$0.03 & 1.85$\pm$0.07 & 9.00$\pm$0.23 & 3.75$\pm$0.50 & 0.75$\pm$0.01 \\
& \cite{zhang2026hierarchical} (open) & 5{,}520$\pm$583 & 0.83$\pm$0.03 & 0.64$\pm$0.00 & 2.15$\pm$0.22 & 8.55$\pm$0.46 & 3.67$\pm$0.56 & 0.79$\pm$0.06 \\
& \cite{zhang2026hierarchical} (close) & 5{,}700$\pm$634 & \underline{0.84}$\pm$\underline{0.02} & 0.65$\pm$0.00 & 2.09$\pm$0.20 & 8.40$\pm$0.51 & 3.52$\pm$0.58 & 0.79$\pm$0.06 \\
& single-skill & 5{,}413$\pm$301 & 0.65$\pm$0.02 & {0.53}$\pm${0.00} & 2.29$\pm$0.11 & 6.54$\pm$0.22 & 1.97$\pm$0.27 & 0.69$\pm$0.01 \\
& RideSkill (w/o reps) & \underline{9{,}144}$\pm$\underline{303} & 0.77$\pm$0.00 & \underline{0.70}$\pm$\underline{0.00} & \textbf{1.10}$\pm$\textbf{0.00} & \textbf{4.58}$\pm$\textbf{0.01} & \textbf{0.22}$\pm$\textbf{0.00} & {0.82}$\pm${0.00} \\
& {RideSkill} & \textbf{9{,}709}$\pm$\textbf{301} & {0.82}$\pm${0.00} & \textbf{0.74}$\pm$\textbf{0.00} & \underline{1.23}$\pm$\underline{0.00} & \underline{4.74}$\pm$\underline{0.00} & \textbf{0.22}$\pm$\textbf{0.00} & \textbf{0.95}$\pm$\textbf{0.00} \\
\bottomrule
\end{tabular}}
\end{table}

\section{Conclusion}

In this paper, we propose RideSkill, which is, to the best of our knowledge, the first ride-sharing algorithm trained via LLM-based evolution and the first that can adapt to varying objectives and operational scenarios. RideSkill is built on a skill repository, from which a combiner chooses a suitable skill for each driver given the objective, the environment scenario, and fairness considerations, followed by a sequential repositioner that balances the distribution of demand and idle vehicles. To adapt the ride-sharing task, we propose an LLM-based ES algorithm with a GDPO-style fitness that balances reward across tasks and a self-check mechanism that aligns the design intention with the realized behavior. To validate our method, we conduct a series of experiments on a real-world ride-hailing dataset in Manhattan, New York, comparing against different types of benchmarks. The results show that our method achieves the best performance across all tasks, with the ability to adapt correspondingly to different tasks. In the future, this approach could be extended to other urban mobility tasks and to settings where the objective varies within an episode. 

\section*{Ethics Statement}
\emph{This work adheres to the principles outlined in the ICLR Code of Ethics.}

\bibliography{refs}
\bibliographystyle{iclr2026_conference}

\appendix
\clearpage
\section*{Appendix Contents}  
\startcontents  
\printcontents{}{1}{\setcounter{tocdepth}{2}}  
\newpage

\section{Related Work} \label{app:literature}
\subsection{Order Dispatch in Ride-Sharing}

Order dispatch in ride-sharing generalizes classical vehicle routing and assignment problems by requiring multiple orders with overlapping itineraries to be bundled onto shared-capacity vehicles. One influential line is the Request-Trip-Vehicle (RTV) framework of \citet{alonso2017demand}, which enumerates feasible order bundles, links them to compatible vehicles with cost-weighted edges, and obtains a cost-minimizing assignment by solving a bipartite matching problem, together with a demand-driven vehicle-rebalancing step. Because constructing and solving such a program is expensive at scale, subsequent model-based methods trade optimality for speed, for example by restricting each vehicle to at most one new request per epoch and relying on implicit bundling of en-route and incoming orders \citep{simonetto2019real}. To counter the myopia of one-shot matching, later works inject future information, such as demand forecasts appended to the assignment graph \citep{alonso2017predictive} and rolling-horizon or model-predictive control for joint relocation and dispatch \citep{riley2021real}. These methods are strong under a fixed operating condition, but they rely on hand-crafted cost models and a single, stationary objective, so they adapt poorly when either the environment or the platform's goal drifts.

A second line learns the dispatch policy from interaction. \citet{xu2018large} first scaled reinforcement learning to ride-hailing by learning a per-vehicle value function and recovering the assignment through global bipartite matching on value-weighted edges. Many subsequent works inherit this learn-a-value-then-match paradigm \citep{qin2020ride,hu2025bmg}, typically under a multi-agent (MARL) formulation that decomposes the large joint dispatch action across vehicles. Decentralized training treats each vehicle as an independent learner \citep{al2019deeppool}, which is simple but suffers from non-stationarity and weak coordination; neighbor-aware encoders such as graph attention \citep{hu2025bmg} and mean-field approximations \citep{li2019efficient} partially mitigate this. Centralized training pursues stronger cooperation through centralized critics and value decomposition, adapting architectures such as hybrid or centralized-critic variants \citep{enders2023hybrid,hoppe2024global} and hierarchical value decomposition \citep{hao2022hierarchical} to the dispatch setting; \citet{jin2025comprehensive} survey the resulting taxonomy. These policies achieve strong performance under the training distribution, but a separate policy must be retrained for each new fleet size, demand regime, or platform objective, and the learned value function transfers poorly across them. RideSkill addresses this by training a single frozen pipeline that adapts zero-shot at inference time, without retraining.

\subsection{LLMs for Ride Hailing}

Several recent works have incorporated LLMs into ride-hailing dispatch, aiming to leverage their generalization capabilities and expert knowledge. For instance, \citet{lyu2026llm} employ an LLM as a per-vehicle decision agent that directly selects orders. Similarly, \citet{zhang2026large} propose a global LLM-based matcher, using the LLM to score vehicles, orders, and vehicle-order pairs for bipartite matching. Both approaches require one LLM call per vehicle per decision step at deployment, making them impractical for fleets exceeding a few hundred vehicles under real-time latency constraints. In a different direction, \citet{su2025data} directly use LLMs to generate the overall dispatch plan based on global input information. Although this method requires only one LLM call per step, the large input and output spaces render it infeasible for real-world large-scale scenarios involving hundreds to thousands of vehicles. An alternative line of work uses LLMs for automatic algorithm design. For example, \citet{zhang2026hierarchical} combine LLM generation with evolutionary search to produce dispatch heuristics. These offline methods eliminate runtime LLM calls. However, all of the aforementioned methods are designed for ride-hailing (single-order dispatch without sharing), require retraining for each new objective, and have not been validated on ride-sharing scenarios with multi-order vehicle loading. Beyond direct dispatch, some works leverage LLMs to enhance ride-hailing tasks without directly using them for order assignment. \citet{lyu2026profillm} propose ProfiLLM, which uses LLMs to analyze passenger and driver profiles, encoding the resulting information as embeddings to improve the final matching model, aiming to reduce cancellation rates on both sides. \citet{jiang2026rideagent} introduce RideAgent for fleet operation in ride-hailing, focusing on region-level operations rather than vehicle-order level assignments, which is a key distinction between ride-hailing and ride-sharing. Table~\ref{tab:comparison} summarizes the key differences between RideSkill and previous LLM-based ride-hailing methods. Notably, RideSkill is the first LLM-based ride-sharing solution, and the first ride-sharing approach capable of adapting to varying scenarios and objectives.

\begin{table}[t!]
\centering
\caption{Comparison of different LLM-based order-dispatch methods: $n$ denotes the number of vehicles and $M$ the number of orders.}
\label{tab:comparison}
\adjustbox{max width=\textwidth}{%
\begin{tabular}{@{} c | c c c c c c c c  @{}}
\toprule
\textbf{Method} & \textbf{Adaptation} & \textbf{\makecell{Generalization\\ \& Transferability}} & \textbf{Sharing} & \textbf{Fairness} & \textbf{Reposition} & \textbf{\makecell{Delay \\ Matching}} & \textbf{Scalability} & \textbf{\makecell{LLM Calls \\ Per Step}}  \\
\midrule
\makecell{LLM-DR\\\citep{lyu2026llm}}   & $\times$ & $\times$ & $\times$ & $\times$ & $\times$  & $\times$ & $\times$ & $n$   \\
\makecell{LLM-ODDR\\\citep{zhang2026large}} & fine-tune  & $\times$ & $\times$ & $\checkmark$ & $\checkmark$ & $\times$ & $\times$  & $O(n{+}M{+}nM)$  \\
\citet{su2025data} & \makecell{fine-tune \\ (optional)} & $\times$ & $\times$ & $\times$ & $\checkmark$  & $\checkmark$ & $\times$ & 1  \\
\citet{zhang2026hierarchical}  & evolution & $\times$ & $\times$ & $\times$ & $\checkmark$ & $\times$ & $\checkmark$ & \makecell{0 (open-loop)\\1 (close-loop)}   \\
\midrule
\makecell{RideSkill\\(ours)}     & evolution & $\checkmark$ & $\checkmark$ & \makecell{$\checkmark$ \\ (optional)} & $\checkmark$ & $\checkmark$ & $\checkmark$ & 0    \\
\bottomrule
\end{tabular}%
}
\end{table}

\section{Notation}
\label{app:notation}

Table~\ref{tab:notation} summarizes the notation used throughout the paper.

\begin{table}[t!]
\centering
\caption{Summary of notation.}
\label{tab:notation}
\smallskip
\small
\adjustbox{max width=\textwidth}{
\begin{tabular}{@{} l l @{}}
\toprule
\textbf{Symbol} & \textbf{Meaning} \\
\midrule
\multicolumn{2}{@{}l}{\emph{Problem formulation}} \\
$\task = \langle n, S, U, \mathcal{P}, \mathrm{R}, \gamma, O, T \rangle$ & MAMDP (a task): fleet size, joint state, joint action, transition, \\
 & reward, discount factor, joint observation, horizon \\
$\Delta t$ & interval between consecutive decision steps \\
$\mathcal{I} = \{1, \ldots, n\}$; $i$ & vehicle (agent) index set; vehicle index \\
$\mathcal{J}_t$; $j$; $\varnothing$ & pending orders at step $t$; order index; dummy no-op order \\
$o_j$; $h_j$ & candidate order $j$; its passenger count \\
$c_i$ & remaining capacity of vehicle $i$ \\
$\obs_i$ & local observation of vehicle $i$ \\
$\mathcal{G}$; $g$ & region set; region index \\
$\ret$ & return (cumulative reward); $\ret^{\mathrm{repo}}, \ret^{\mathrm{off}}$: paired returns with repositioning on/off \\
\midrule
\multicolumn{2}{@{}l}{\emph{RideSkill components}} \\
$\mathcal{B} = \{s_1, \ldots, s_K\}$; $K$ & skill repository; its size \\
$s_k(\obs_i, o_j, \phiep, \phistep)$ & score of skill $k$ for the vehicle-order pair $(i,j)$ \\
$\phiep$; $\phistep$ & episode-static context; live per-step context \\
$\kappa = (\kappa_g)_{g \in \mathcal{G}}$ & shared per-region effective demand/supply state \\
$w$ & platform objective: a per-step reward function on event descriptors \\
$\pi_c$; $\pi_r$ & combiner; repositioner \\
$\alpha_i \in \mathbb{R}^K$; $\tilde{\alpha}_{i,k}$ & raw skill scores of vehicle $i$; softmax weights over $\mathcal{K}_i$ \\
$b$; $\mathcal{K}_i$ & number of blended skills; the top-$b$ skill set of vehicle $i$ \\
$\mathcal{J}_i$; $\mu_k, \sigma_k$ & feasible candidates of vehicle $i$; mean/std of $s_k$ over $\mathcal{J}_i$ \\
$a_{i,j}$; $u_{i,j}$ & matching score; assignment variable of pair $(i,j)$ \\
$\mathcal{G}_i$; $\nu_g$; $H$ & candidate regions of vehicle $i$; relocation score; hottest-region count \\
$E_i$; $\bar{E}, \sigma_E$ & cumulative income of vehicle $i$; its fleet-wide mean and std \\
$z_i$; $\beta_i$; $\rho$; $\epsilon$ & income z-score; fairness budget; fairness strength; small constant \\
\midrule
\multicolumn{2}{@{}l}{\emph{Training}} \\
$\mathcal{T} = \{\task_m\}$; $m$ & task batch; task index, with reward $\mathrm{R}_m$ and dynamics $\mathcal{P}_m$ \\
$\pi_l$; $l$ & candidate policy; candidate index \\
$\ret_l(\task_m)$ & return of candidate $\pi_l$ on task $\task_m$ \\
$A_{l,m}$; $\mu^{(m)}, \sigma^{(m)}$ & group-relative advantage; group mean/std on task $\task_m$ \\
$F_l$ & fitness of candidate $\pi_l$ \\
$\mu$, $\lambda$; $p_{\mathrm{cross}}$ & ES parent/child population sizes; crossover probability \\
$N_{\mathrm{gen}}$, $N_{\min}$, $N_{\mathrm{pat}}$ & generation cap; minimum generations; patience \\
$N_f$, $N_a$, $N_r$ & distinct-failure cap; audit retry cap; repair cap \\
$\tau$ & behavioral-dedup cosine-similarity threshold \\
$w_{\mathrm{det}}, w_{\mathrm{pick}}, w_{\mathrm{serv}}$ & the objective's coefficients on detour, pickup, and service time \\
\bottomrule
\end{tabular}
}
\end{table}

\paragraph{Terminology.} To avoid confusion, we define the terminology used in this paper formally.
\begin{itemize}[left=0pt]
\item A \textbf{scenario} is a specific environmental configuration: fleet size, vehicle capacity, speed, and a demand window drawn from historical trip records. (related to $n, S, U, \mathcal{P}, O, T$)
\item An \textbf{objective} is the platform's optimization goal, encoded as a reward function $\mathrm{R}$ that assigns a scalar reward to each vehicle per step. (related to $\mathrm{R}, \gamma$)
\item A \textbf{task} is the pairing of one scenario and one objective. (related to the whole $\task$)
\item A \textbf{policy} (or program in the context of LLM automatic algorithm design) refers to a complete dispatch solution (dispatcher, repositioner, or both). (referred to as $\pi$)
\end{itemize}

\section{Method Implementation} \label{app:imp}

\subsection{Prompt Design}
\subsubsection{Skill Creation (Phase 1)}
\label{app:prompt-skill}

Phase 1 evolves one objective-specialist scoring function at a time. The model first declares the objective it will specialize in, then writes a self-authored fitness function over the metric menu (which remains fixed throughout the search). To encourage diversity, the system provides the model with descriptions of skills already present in the basis, and uses a mechanism menu that enforces a genuinely different decision-rule shape.

\begin{lstlisting}[style=promptbox, caption={System prompt: Phase-1 skill creation.}]
You are an expert in ride-POOLING dispatch and reward design. You write small,
robust, interpretable Python scoring functions for a fleet dispatcher. You reason
carefully about the objective, then produce clean code AND clear natural-language
explanations of the dispatch behaviour. You always answer with exactly one JSON
object matching the requested schema.
\end{lstlisting}

\begin{lstlisting}[style=promptbox, caption={Output contract: one skill proposal. Explanation fields come first and are mandatory.}]
Respond with ONE JSON object and nothing else (no prose before/after, no markdown
outside the JSON). Schema:

{
  "skill_name": "<short snake_case id, e.g. long_fare_hunter>",
  "objective":  "<ONE sentence: the single objective this skill specialises in>",
  "objective_self_check": "<2-3 sentences: WHICH objective axis this skill covers and
                 whether that axis is already covered by the existing skills. Name
                 the axis (see the OBJECTIVE AXES list); say which listed skill (if
                 any) is close, and what is actually NOT yet covered that this skill
                 is covering. If the axis IS already covered, justify why a second
                 specialist on it is still worth a skill slot.>",
  "mechanism":  "<ONE phrase naming the DECISION RULE SHAPE you used, e.g.
                 'hard feasibility gate then fare-per-minute ranking',
                 'idle-time-triggered threshold switch', 'ratio of marginal fare to
                 marginal detour', 'two-stage: shortlist by pickup, break ties by
                 pooling slack'>",
  "differs_from":"<1-2 sentences: which listed skill/mechanism yours is closest to
                 and what it does DIFFERENTLY at decision time -- not 'different
                 weights', an actually different rule>",
  "description":"<2-4 sentences: HOW the score logic realises that objective and
                 when it prefers to wait; explain behaviour, not code>",
  "fitness_code": "def fitness(metrics):\n    # cheap scalar over the metrics dict\n    return ...",
  "fitness_rationale": "<1-2 sentences: why this fitness measures the objective>",
  "code": "def score(driver_obs, order, phi_ep, phi_step):\n    ...\n\ndef noop_score(driver_obs, phi_ep, phi_step):\n    ..."
}

Rules for the fitness function:
  - It is YOUR self-authored reward for THIS skill; it need not be comparable to
    any other skill's fitness. It only has to rank this skill's own variants.
  - It must be a cheap pure function of the metrics dict only (arithmetic over the
    keys listed in the metrics menu). No rollouts, no env, no randomness, no LLM calls.
  - Same sandbox rules as the skill code (no imports; math/np only).

SELF-CHECK BEFORE YOU SUBMIT (do this EVERY time you write a NEW skill):
After writing `code`, confirm BOTH:
  (a) AXIS -- your objective specialises on ONE axis from the OBJECTIVE AXES list
      (a genuinely different objective is a different customer, not a reworded one).
  (b) COVERAGE -- either your axis is NOT covered by any listed skill, or you state
      in `objective_self_check` why a second specialist on an already-covered axis
      is still a distinct behaviour (a different decision rule genuinely aimed at
      that same axis -- never just different weights). A repository whose skills
      all optimise the same two axes leaves the other axes UNANSWERABLE, and the
      upper combiner cannot then respond to a reward pricing one of them.
      Axes: revenue / fare, service / wait, throughput, detour on a new order,
      remaining capacity, empty / idle cost, fairness, option value (patience).
\end{lstlisting}

\begin{lstlisting}[style=promptbox, caption={Mechanism menu: behaviourally distinct decision-rule shapes (skill creation).}]
The `mechanism` field must name the SHAPE of the decision rule, and the shapes
below are all legitimate and behaviourally distinct. Pick one that the existing
repository does not already use, or invent another:

  - weighted sum of terms (the default -- already well covered, prefer something else)
  - ratio / efficiency (value per unit of a cost: fare per minute, revenue per km)
  - hard gate then rank (reject anything failing a condition, rank only survivors)
  - threshold switch on a state variable (behave one way when idle_min > T or
    demand_pressure > P, a different way otherwise)
  - two-stage / lexicographic (shortlist by one criterion, break ties by another)
  - marginal / counterfactual (score the CHANGE this order causes: added detour for
    the passengers already on board, capacity consumed, time-window slack burnt)
  - opportunity cost (what accepting this order costs you in orders you can no
    longer reach -- compare against noop_score deliberately)
  - patience / option value (a high noop_score that makes waiting a real choice, so
    the driver holds out for a better match instead of taking the first feasible one)
  - non-linear saturation (diminishing returns above a value, cliff below a value)

A skill whose mechanism is "weighted sum" with new coefficients is NOT a new skill.
\end{lstlisting}

\begin{lstlisting}[style=promptbox, caption={Available metrics: the exact dict keys a self-authored fitness may read, with direction and typical one-hour ranges.}]
An episode rollout returns this metrics dict (these EXACT keys; a fitness may
only read from here). Direction = what "better" means for that term. TYPICAL
RANGE is measured over one real hour at the two ends of the fleet range
(200 cars .. 1800 cars):

  revenue            float  higher better  -- sum over assigned orders of
                                              solo_time(min) x party size.
                                              TYPICAL 36,000 .. 107,000
  service_rate       float  higher better  -- assigned / total_orders, in [0,1].
                                              TYPICAL 0.20 .. 0.98
  completed          int    higher better  -- orders actually delivered.
                                              TYPICAL 1,200 .. 6,200
  assigned           int    higher better  -- orders assigned to a driver.
                                              TYPICAL 1,800 .. 8,400
  mean_service_time  float  lower  better  -- mean end-to-end service time (min).
                                              TYPICAL 11 .. 18
  detour_total       float  lower  better  -- total extra detour time from
                                              pooling (min); the pooling cost.
                                              TYPICAL 10,000 .. 38,000
  income_gini        float  lower  better  -- driver-income inequality, [0,1].
                                              TYPICAL 0.15 .. 0.17
  income_cv          float  lower  better  -- driver-income coeff. of variation.
                                              TYPICAL 0.77 .. 0.94
  income_mean        float  (context)      -- mean per-driver cumulative reward.
                                              TYPICAL 2.3 .. 3.1
  income_min         float  higher better  -- worst-off driver's cumulative
                                              reward. TYPICAL -5.5 .. -4.8.

MECHANICS. The fitness you write at generation 0 is FIXED for the whole search:
every later variant of this skill is graded by it. Before you submit, put the
TYPICAL numbers above into your own formula and check the ordering it gives.
\end{lstlisting}

\subsubsection{Combiner Creation (Phase 2)}
\label{app:prompt-combiner}

Phase 2 evolves the upper combiner $\pi_c$ on top of the frozen skill basis. Unlike Phase 1, the combiner does not author a fitness function; that is fixed by the researcher. Instead, it implements an objective-reading function, \lstinline{skill_scores(driver_obs, phi_ep, phi_step, w)}, which, for each vehicle, scores every frozen skill so that the resulting blend serves the episode objective $w$, which it has not seen during training. Since the reward is an opaque function that the combiner must infer by probing, the contract includes a probe-event evolution specification. The model is informed that $w$ is a per-step linear price vector, and is provided with the exact event-dictionary keys. It is required to construct synthetic probe events, each isolating one term (e.g., completion, seating/party volume, solo length, dispatch wait, pickup time, detour on a new order versus detour on onboard orders), and to invoke $w$ on these events to read the corresponding coefficient as a difference. When the objective is supplied as a concrete reward function, a mandatory \lstinline{reward_understanding} chain-of-thought field is prepended to the contract.

\begin{lstlisting}[style=promptbox, caption={System prompt: Phase-2 combiner creation.}]
You are an expert in ride-POOLING fleet dispatch and objective-conditioned policy
design. A set of lower-layer scoring skills is ALREADY FROZEN; you cannot change or
add skills -- you only decide, per driver and per episode objective, WHICH frozen
skill that driver should use. You write one small, robust, interpretable Python
function AND a clear natural-language explanation of the policy. You always answer
with exactly one JSON object matching the schema.
\end{lstlisting}

\begin{lstlisting}[style=promptbox, caption={Output contract: one combiner (reward-conditioned variant; the base variant omits the first field).}]
Respond with ONE JSON object and nothing else (no prose before/after, no markdown
outside the JSON). Schema:

{
  "reward_understanding": "<2-4 sentences, FIRST: in your OWN words, what the
                    REWARD FUNCTION above rewards and penalises, and what a
                    reward-maximising dispatcher must therefore do. This is a
                    chain-of-thought gate: reason about the objective BEFORE you
                    compose the skills.>",
  "combiner_name": "<short snake_case id, e.g. reward_aware_dispatcher>",
  "strategy":      "<ONE sentence: how you turn that reward + driver state into a
                    skill choice>",
  "description":   "<3-5 sentences: which driver STATES you distinguish (idle /
                    loaded-with-slack / deadline-pressed / ...), which skill each
                    tends to get, and HOW the reward's terms (throughput / revenue /
                    service / detour) drive those choices. Explain behaviour, not code.>",
  "probe_self_check": "<2-4 sentences: CONFIRM your probe set is diverse enough. List
                    the terms it differentiates (dispatch_wait, pickup_time,
                    solo/service time, detour on a NEW order, detour on ONBOARD
                    orders, completion, seating/party, volume, empty-move, idle-wait)
                    and note any extra probes you added beyond that list. If you only
                    probe two or three terms, say so -- a shallow probe set is a
                    failed combiner.>",
  "code": "def skill_scores(driver_obs, phi_ep, phi_step, w):\n    ...\n    return {<skill>: <score>, ...}"
}

SELF-CHECK BEFORE YOU SUBMIT (every time you write a combiner):
  (a) COVERAGE -- each of these KNOWN terms is the differentiating factor in at
      least one probe (or you explicitly justify dropping it in
      `probe_self_check`): dispatch_wait, pickup_time, solo_time/service_time,
      detour on a NEW order, detour on ONBOARD orders, completion, seating/party
      size, volume, empty-move, idle-wait. A reward may price ANY of these, and
      the same frozen combiner must read ANY future reward.
  (b) EXPLORE -- add probes for terms the reward MIGHT price beyond that list.
      Extra probes are HARMLESS: the combiner is not required to use every probe,
      and a probe this reward ignores may be the ONLY thing that reveals a penalty
      in the NEXT reward. If you probe only two or three terms, add more.
\end{lstlisting}

\begin{lstlisting}[style=promptbox, caption={Function contract: how the scores are consumed by the platform (blending rule).}]
Your function MUST have this exact signature (do not change it):

    def skill_scores(driver_obs, phi_ep, phi_step, w) -> dict:
        # return {skill_name: score} scoring the FROZEN skills for THIS driver.

HOW YOUR SCORES ARE USED. The platform takes your {skill}->{score} dict, keeps the
B highest-scoring skills with a positive score, softmax-normalises those scores
into weights, standardises each skill across this driver's candidate orders, and
dispatches on the weighted sum. So:
  - your RELATIVE scores matter, not just which one is largest;
  - scoring exactly one skill and zeroing the rest is legal but throws away the blend;
  - the blend is per DRIVER per STEP, so different cars can carry different mixes.

Rules:
  - Only use the frozen skill names listed as keys; any other key is rejected.
  - Score at least one known skill for every driver.
  - Distinguish drivers by STATE; read the objective through `w` and the live scene
    through `phi_step`; never hard-code a fixed operating point.
  - Handle `w is None` gracefully; handle a driver with no nearby orders gracefully.
\end{lstlisting}

\begin{lstlisting}[style=promptbox, caption={Probe-event evolution specification (abridged): how to build the probe events and read a coefficient.}]
The objective `w` is called on a small event dict; read its per-term price as a
DIFFERENCE between two events that differ only in that term. For example:
  detour_onboard_signal = w(bundling_with_detour) - w(bundling_without_detour)
  detour_new_signal     = w(new_order_with_detour) - w(new_order_without_detour)
  dispatch_signal       = w(dispatch_wait_nonzero) - w(dispatch_wait_zero)
  pickup_signal         = w(long_pickup) - w(short_pickup)
  completion_signal     = w(completion) - w(no_completion)
  seating_signal        = w(party_2) - w(party_1)
These differences give the reward's coefficient per term -- and, for detour, the
SEPARATE coefficient on new-order detour vs. onboard detour.

MANDATORY PROBE RULES: create at least one probe where dispatch_wait is non-zero
and one where it is zero; one where pickup_time differs substantially from
solo_time; one with completed_orders non-empty and one with it empty; one with
party_size > 1 and one with party_size = 1; and distinguish NEW orders (in
assigned_orders) from ALREADY-CARRIED orders (in picked_up_orders), because the
reward may price the two groups differently.

DESIGN PRINCIPLES: every reward term must be the differentiating factor in at
least one probe; probe count can be GENEROUS -- an extra probe costs you nothing
and buys robustness across objectives; use realistic magnitudes (solo_time,
detour, pickup_wait, party_size); both onboard-order and new-order terms must be
represented.
\end{lstlisting}

\begin{lstlisting}[style=promptbox, caption={The event dict the combiner probes (a linear price list event).}]
An event dict to probe `w` with has keys:
  assigned_orders (list), assigned_party_sizes (dict),
  assigned_dispatch_wait (dict), assigned_pickup_times (dict),
  assigned_solo_times (dict), assigned_service_times (dict),
  assigned_detour_times (dict), completed_orders (list),
  picked_up_orders (list), distance_moved (float), time_moved (float),
  is_empty_move (bool), is_idle_wait (bool), extra_detour_time (float).
`assigned_detour_times[oid]` is the per-order pooled detour on a NEW order;
`extra_detour_time` is the SIGNED aggregate re-routing impact on ONBOARD orders.
\end{lstlisting}

\subsubsection{Repositioner Creation (Phase 3)}
\label{app:prompt-repositioner}

Phase 3 evolves the per-region repositioning scorer. The objective and fitness are fixed by the researcher: the scorer is evaluated solely based on the delta $\ret^{\mathrm{repo}} - \ret^{\mathrm{off}}$ (repositioning on versus off, on the same random seed), standardized within each round's group, so the only target is to outperform the "do nothing" baseline. The model is responsible only for writing the base per-region score; the spreading logic, stay rules, and the emitted action are handled outside the model. The contract requires the scorer to state, in a \lstinline{objective_read_check} field, how the target region depends on the objective $w$. A program that reads $w$ only to compute a score but never lets it influence the argmax is considered purely cosmetic; this is detected and reported by the round's objective-blindness metric.

\begin{lstlisting}[style=promptbox, caption={System prompt: Phase-3 reposition-scorer creation.}]
You are an expert in ride-POOLING fleet operations and empty-vehicle
repositioning. You are given a FIXED objective (send idle empty cars toward
near-future demand without wasteful cruising) and a FIXED fitness (the service
improvement it brings over not repositioning). You must first explain, in natural
language, what makes a good reposition target, and then write ONE small, robust,
interpretable Python scorer that rates each preset region for an idle driver. You
always answer with exactly one JSON object matching the requested schema.
\end{lstlisting}

\begin{lstlisting}[style=promptbox, caption={Output contract: one reposition scorer (no fitness field; the fitness is fixed and researcher-given).}]
Respond with ONE JSON object and nothing else (no prose before/after, no markdown
outside the JSON). Schema:

{
  "reposition_understanding": "<2-4 sentences, FIRST: in your own words, what makes
                 a region worth cruising an idle empty car toward, and what a
                 service-maximising, waste-avoiding reposition scorer must therefore
                 do. This is a chain-of-thought gate: reason before you write.>",
  "skill_name":  "<short snake_case id, e.g. demand_gravity_scorer>",
  "objective":   "<ONE sentence: the per-region scoring policy you will use>",
  "objective_read_check": "<2-3 sentences: HOW your scorer makes the TARGET REGION
                 depend on the objective `w` (not just the score magnitude). Name at
                 least one objective family you respond to and the concrete mechanism
                 -- e.g. a completion-gated w pushes you to cruise only to regions
                 with genuinely imminent pickups, an empty-averse w pushes you away
                 from long empty cruises, a seating w pushes you toward multi-party
                 regions, a length-driven w toward long-fare origins. If the described
                 mechanism does not change WHICH region wins the argmax when `w`
                 changes, it is cosmetic and will be caught by the blindness report.>",
  "description": "<2-4 sentences: HOW your scores rank regions -- how you weigh
                 demand, cruise distance, and existing supply, and when you return
                 {} / a low score to keep a car put. Explain behaviour, not code.>",
  "code": "def reposition_scores(driver_obs, phi_ep, phi_step, kappa, w):\n    ..."
}

There is NO fitness field: the objective and fitness are fixed and given above; you
only write the scoring policy that maximises the fixed fitness.
\end{lstlisting}

\begin{lstlisting}[style=promptbox, caption={Fixed objective and delta fitness the scorer is graded by (abridged).}]
You ONLY provide the per-region base score; the dispatcher applies its own
deterministic logic on top (spreading, stay rules, the actual relocate action).

[...] You are NOT scored on reward directly -- you are scored on WHAT YOUR
REPOSITIONING IS WORTH. On every (scene, objective, fairness-strength) cell:

    (YOUR episode reward - the episode reward with repositioning switched OFF,
     i.e. every idle car left parked, on the SAME scene and the SAME seed)
    / how much this round's programs disagree about that same difference

[...] THE SIGN IS ABSOLUTE, not a rank. 0.00 means your scorer was worth exactly
as much as leaving every car parked; NEGATIVE means sending cars around ACTIVELY
lost money; +1 means one full spread above the field. Beating the other candidates
is NOT the target -- beating "do nothing" is. The objective `w` VARIES across
episodes (completion-gated, pooling/seating, throughput, empty-averse families;
all LINEAR in the per-step event terms); read `w` AND the live kappa together.
\end{lstlisting}

\subsubsection{Reward Authoring (pre-Phase 2, NL to code)}
\label{app:prompt-reward}

Before the combiner is composed, a platform preference—expressed either as a natural-language description or a weight vector—is translated into a concrete per-vehicle, per-step reward function \lstinline{reward(event) -> float}. This reward function is injected directly into the environment, and its fleet-average cumulative value becomes the fitness that the combiner is later optimized to maximize.
After the reward is authored, its coefficients are normalized so that they sum to one. Specifically, the source code is parsed to extract the named constant prices, and the entire function is divided by their sum. This is a single global rescaling operation that preserves the exact ratio between coefficients as intended by the author, and it is applied to the reward function itself—not to the objective $w$ that the combiner later probes.
The combiner reads $w$ by probing differences between events, and any single global scale cancels out in these differences. Therefore, normalizing the coefficients (rather than the $w$ responses) ensures that the term ratios read by the combiner remain exactly those specified by the author.

\begin{lstlisting}[style=promptbox, caption={System prompt: reward authoring.}]
You are an expert in ride-POOLING fleet economics and reinforcement-learning
reward design. You are given a platform PREFERENCE (natural language or weights)
and the exact per-step event signals available. You must first explain, in
natural language, what that preference wants, and then write ONE small, robust,
interpretable Python reward function `reward(event) -> float` that encodes it. You
always answer with exactly one JSON object matching the requested schema.
\end{lstlisting}

\begin{lstlisting}[style=promptbox, caption={Output contract: one authored reward function.}]
Respond with ONE JSON object and nothing else (no prose before/after, no markdown
outside the JSON). Schema:

{
  "reward_understanding": "<2-4 sentences, FIRST: in your own words, what the
                 platform PREFERENCE below wants, and which quantities in the
                 per-step event dict must therefore be rewarded or penalised (and
                 roughly how strongly). If the preference asks for something the
                 additivity rule forbids (a threshold, a ratio, an escalating
                 bonus).>",
  "reward_name": "<short snake_case id, e.g. completion_first_reward>",
  "objective":  "<ONE sentence: what a driver maximising this reward will do>",
  "description": "<2-4 sentences: HOW each term of your reward encodes the
                 preference -- what it pays for, what it charges, and the relative
                 magnitudes. Explain behaviour, not code.>",
  "code": "def reward(event):\n    ..."
}

Write each per-order/per-step price as a NAMED CONSTANT (e.g. `WAIT_PRICE = 2.0`)
that is used in a product with an `event[...]` field, so the authoring stage can
recover the coefficient list and normalise it.
\end{lstlisting}




\subsubsection{Self-Check: Skill Audit (post Phase-1 search)}
\label{app:prompt-skill-audit}

After a Phase-1 search completes, the champion is asked to evaluate its own output: given the measured rollout table, it decides whether the observed behavior matches the behavior it originally intended to implement. The resulting judgment falls into one of three categories, which map to two distinct repair strategies: either rewriting the program description, or re-authoring the fitness function and restarting the search. This audit is advisory only; its verdict does not influence the fitness value.

\begin{lstlisting}[style=promptbox, caption={System prompt: Phase-1 skill audit.}]
You are auditing a dispatch skill that you previously designed and evolved. You
are shown what you SAID the skill would do and what it MEASURABLY did. You are
blunt and evidence-driven: you quote the numbers that decide the verdict, you do
not defend the design, and you do not invent behaviour the table does not show.
You always answer with exactly one JSON object matching the requested schema.
\end{lstlisting}

\begin{lstlisting}[style=promptbox, caption={Output contract and verdict choices: skill audit.}]
{
  "verdict": "match" | "description_wrong" | "fitness_wrong",
  "reason":  "<2-4 sentences: what the measured behaviour actually IS, and how it
              relates to what was promised. Quote the specific numbers.>",
  "evidence":"<ONE sentence naming the columns/rows that decided it>",
  "new_objective":    "<REQUIRED iff verdict is description_wrong: one sentence,
                       the objective this skill actually pursues>",
  "new_description":  "<REQUIRED iff verdict is description_wrong: 2-4 sentences
                       describing the behaviour in the table>",
  "fitness_complaint":"<REQUIRED iff verdict is fitness_wrong: what the fitness
                       formula actually pays for, which term dominates it at the
                       magnitudes in the table, and what a skill maximising it is
                       therefore driven to do. Be concrete and arithmetic.>"
}

Choosing the verdict:
  "match"              The table is a recognisable instance of the intended
                       behaviour (a weak specialist in the right thing counts).
                       MECHANICALLY: the skill is frozen as-is.
  "description_wrong"  The skill does something coherent, but not what the
                       description claims; the fitness rewards what the table shows.
                       MECHANICALLY: new_objective / new_description replace the old
                       ones and the skill is frozen; no re-search.
  "fitness_wrong"      The fitness's own maximum sits somewhere other than the
                       intended behaviour. MECHANICALLY: the champion is discarded,
                       the fitness re-authored from fitness_complaint, and the whole
                       search re-run. This is the only verdict that costs a re-search,
                       and re-searches are capped.
\end{lstlisting}

\subsubsection{Policy Audit: Soft Check (post Phase-2/3 search)}
\label{app:prompt-policy-audit}

Phases 2 and 3 have fixed fitness functions, so they are not subject to the incorrect-fitness failure that can occur in Phase 1. However, they can suffer from a different shortcoming—one that the overall method aims to avoid: the combiner or repositioner may entirely ignore the objective $w$. To detect this, the audit performs a counterfactual measurement: the same demand hour is rolled out twice with the same random seed, differing only in whether the program receives $w$ as input. The resulting verdict is recorded alongside the frozen artifact and later reviewed manually.

\begin{lstlisting}[style=promptbox, caption={System prompt: Phase-2/3 policy audit (objective-responsiveness soft check).}]
You are auditing a dispatch program that you previously designed and evolved. You
are shown what you SAID it would do and a set of paired rollouts that isolate one
thing: what changed when the program was handed the episode objective instead of
being run without it. You are blunt and evidence-driven: you quote the numbers that
decide the verdict, you do not defend the design, and you do not invent behaviour
the table does not show. You always answer with exactly one JSON object matching
the requested schema.
\end{lstlisting}

\begin{lstlisting}[style=promptbox, caption={Output contract and verdict choices: policy audit.}]
{
  "per_cell": [
     {"cell": <1-based cell number>,
      "moved": "yes" | "no" | "unclear",
      "note": "<ONE sentence: which columns separate the arms in this cell, by how
                much, and how that relates to what THIS cell's objective pays for>"}
     ... one entry per cell, in order ...
  ],
  "verdict": "reads_it" | "moves_elsewhere" | "no_movement" | "mixed",
  "reason":  "<2-4 sentences summarising the per-cell readings. Quote specific
              numbers.>",
  "evidence":"<ONE sentence naming the cells and columns that decided it>"
}

The verdicts:
  "reads_it"        Across the cells, the arm given the objective behaves
                    differently from the arm that was not, and the differences
                    follow from what those objectives pay for.
  "moves_elsewhere" The arms do differ, but the differences do not follow from the
                    price lists shown -- the program changes behaviour on something
                    other than what the objective asks for.
  "no_movement"     The arms are the same to within rounding: handing over the
                    objective changed nothing measurable.
  "mixed"           The cells do not agree with each other.

MECHANICALLY: nothing is applied from this answer. The program is already frozen;
the verdict is recorded next to the artifact and read by a person.
\end{lstlisting}

\subsubsection{Evolution Prompts: Mutation and Crossover}
\label{app:prompt-evolution}

The $(\mu{+}\lambda)$ loop uses two LLM operators that share the same output contract as creation but differ in their task framing.

\paragraph{Mutation (improvement).} When improving a parent, the parent's code and measured fitness are included so the model rewrites to earn a higher score:

\begin{lstlisting}[style=promptbox, caption={Mutation task framing (excerpts from combiner evolution).}]
# TASK
Improve the UPPER combiner. Rewrite skill_scores to earn MORE cumulative
value under the episode objective (below), keeping the same contract.
It must still serve unseen objectives with no retraining.

# PARENT (current best)
<name>, score: <fitness_note>
```python
<parent_code>
```

# SCENE VARIABILITY
THE DEPLOYMENT SCENE ALSO CHANGES (write a SCALE-INVARIANT combiner).
At test time the SAME frozen function is run zero-shot on scenes that
differ in fleet size (~100 to ~2000), capacity (~1 to ~10), speed, and
demand. Do NOT hard-code absolute numbers; express thresholds RELATIVELY
using phi_step.mean_solo_time as the time yardstick.

# OBJECTIVE DIVERSITY
THE STRATEGY MUST CHANGE WITH THE OBJECTIVE FUNCTION. You are writing
a policy that SPECIALISES PER OBJECTIVE. Probe the event SHAPE:
vary whether the probe carries completed orders, party size, service
time, detour, and empty/idle flags, then route drivers to whichever
frozen skill best pursues the dominant coefficient.
\end{lstlisting}

During mutation, the model receives the parent's complete code, its measured fitness (including a per-family breakdown and the objective-blindness metric), and any repair feedback from previous failed attempts. It must preserve the same \lstinline{skill_scores} contract while modifying the dispatch logic to improve upon the parent's identified shortcomings.

\paragraph{Crossover (recombination).} When two parents are recombined, both are presented with their fitness notes, and the model is instructed to merge complementary strengths:

\begin{lstlisting}[style=promptbox, caption={Crossover task framing.}]
# TASK
Design a CHILD combiner by RECOMBINING the two parent programs below.
Both parents survived selection, and each is stronger on a DIFFERENT set
of objective families (their per-family advantages are printed with them).
Your job is not to pick a winner and tweak it: read what each parent
actually does WELL, and build one program that keeps both strengths.

Concretely:
  - Name, for yourself, the mechanism in parent A that wins A's strong
    families and the mechanism in parent B that wins B's strong families.
  - Write ONE skill_scores that routes to the A-mechanism on the
    situations A is good at and the B-mechanism on the ones B is good
    at -- decided by the objective w and the driver's own state.
  - Where the parents disagree on the SAME situation, keep the one
    whose family advantage is higher there.
  - You may add a small improvement, but the child must visibly
    inherit from both parents.

# PARENT A: <name>
strategy: <strategy_summary>
score: <fitness_note>
```python
<parent_A_code>
```

# PARENT B: <name>
strategy: <strategy_summary>
score: <fitness_note>
```python
<parent_B_code>
```
\end{lstlisting}

Each parent is presented with its name, strategy summary, per-family fitness note, and the full code block. The crossover operator ensures that the mechanism of a family specialist can be transferred to a strong all-rounder through recombination, rather than being lost alongside its specialist parent.

\subsubsection{Shared Context Specification}
\label{app:shared-context}

Every layer that receives context objects—including skills, the combiner, and the repositioner—shares the same two-layer specification defined below. This ensures that the field list remains consistent across all prompts. Here, $\phi_{\mathrm{ep}}$ denotes episode-static information, $\phi_{\step}$ represents live per-step context, and \lstinline{driver_obs} and \lstinline{order} are dictionaries accessed via \lstinline{d['key']}. This specification block is defined once in \lstinline{common.py} and included in every prompt to guarantee uniformity.

\begin{lstlisting}[style=promptbox, caption={Shared two-layer context and argument contract.}]
  phi_ep: episode-STATIC context (same object every step).
       An OBJECT, not a dict -- read with attributes.
    dist(a, b)       -> travel time in MINUTES between two (lon,lat) points.
    scale            -> leak-free static map scale (minutes).
    num_drivers      -> fleet size (fixed for the episode).
    driver_capacity  -> per-vehicle seat count (fixed).
    speed_kmh        -> driver speed (fixed).
    region_centres   -> tuple of (lon,lat), one per region, indexed by region id.
    region_neighbours -> tuple of tuples; region_neighbours[i] lists adjacent regions.
    od_count         -> previous-hour OD flow matrix (sums to 1.0).
    od_out[i]        -> share of last-hour orders STARTING in region i.
    od_in[i]         -> share of last-hour orders ENDING in region i.
    od_orders        -> raw order count of last hour (0 if unavailable).

  phi_step: LIVE per-step context (recomputed every step).
       An OBJECT, not a dict -- read with attributes.
    time, num_pending, num_idle, total_free_capacity,
    demand_pressure = pending / total_free_capacity,
    mean_solo_time   -> LIVE SCALE in minutes (use as unit; fall back to
                        phi_ep.scale when ~0 because no orders are pending).
    region_demand[i] / region_supply[i] -> live per-region counts.

  driver_obs["self"]:  DICT (use d['key'] or d.get(k, default))
    location             -> (lon, lat)
    current_region       -> int region index
    status               -> str (idle / to_pickup / to_dropoff / relocating)
    capacity             -> int seat count
    committed_passengers -> int currently onboard
    assigned_order_details -> list of dicts:
        order_id, origin=(lon,lat), destination=(lon,lat),
        num_passengers, onboard(bool), eta(minutes)

  driver_obs["pending_orders"]: list of candidate order dicts:
    order_id, origin=(lon,lat), destination=(lon,lat),
    origin_region(int), destination_region(int),
    num_passengers, waiting_time(minutes)

  driver_obs["relocation_points"]: tuple of (lon,lat) region centres.
  driver_obs["region_neighbours"]: tuple of tuples (adjacency graph).
  driver_obs["fairness_budget"]: float (this driver's multiplier; >1 = boosted).
  driver_obs["driver_budgets"]: dict {driver_id: multiplier} for whole fleet.

  order:  DICT (use d['key'] or d.get(k, default))
    order_id, origin=(lon,lat), destination=(lon,lat),
    origin_region(int), destination_region(int),
    num_passengers, waiting_time(minutes).

KEY RULES:
  driver_obs and order are DICTS -> use d['key'] or d.get(k, default).
  phi_ep and phi_step are OBJECTS -> use phi_ep.scale, phi_step.mean_solo_time.
  region_centres, region_neighbours, od_count, region_demand, region_supply
  are EMPTY tuples when the env has no region layout.
  Test region_id >= 0 before indexing; test phi_ep.od_orders before reading OD.
  Skills do NOT see w; only combiner and repositioner do.
\end{lstlisting}

\begin{lstlisting}[style=promptbox, caption={Shared sandbox rules (identical for skill, combiner, and repositioner code).}]
Code rules (enforced by an AST sandbox -- violating them rejects your program):
  - Pure functions of the given arguments. No import statements of any kind.
  - Allowed globals: math, np (numpy). Allowed builtins: abs, min, max, sum, len,
    float, int, round, sorted, range, enumerate, zip, map, filter, pow, all, any,
    bool, list, dict, tuple, set.
  - No eval/exec/open/getattr and no attribute access beginning with underscore.
  - Never hard-code a distance/time constant: express thresholds in units of
    phi_step.mean_solo_time (the live scale; fall back to phi_ep.scale when ~0).
  - Always return a finite float (skills use -1e9 for infeasible orders).
\end{lstlisting}

\subsection{Algorithm Process}
The training and inference processes are shown as Algorithm \ref{alg:phase1} to Algorithm \ref{alg:runtime}.

\begin{algorithm}[t!]
\caption{Phase-1 Skill Evolution: QD proposal, $(\mu{+}\lambda)$-ES, dedup, and self-check.}
\label{alg:phase1}
\begin{algorithmic}[1]
\Require Environment profile, metric menu, mechanism menu; repository cap $K$; distinct-failure cap $N_f$; audit retry cap $N_a$; repair cap $N_r$; ES parameters $\mu,\lambda$; generation cap $N_{\mathrm{gen}}$; min generations $N_{\min}$; patience $N_{\mathrm{pat}}$; dedup threshold $\tau$.
\Ensure Frozen skill repository $\mathcal{B}=\{s_1,\dots,s_K\}$ with interpretability cards.
\State $\mathcal{B}\leftarrow \{\}$; $\text{fail}\leftarrow 0$
\While{$|\mathcal{B}|<K$ \textbf{and} $\text{fail}<N_f$}
    \State \textbf{Propose.} LLM generates a proposal $\pi_0=(\text{objective},\text{mechanism},\text{fitness},\text{code})$ conditioned on the diversity cards of $\mathcal{B}$ and the mechanism menu; it must additionally state, in \lstinline{objective_self_check}, which objective axis it covers and that the axis is not already saturated in $\mathcal{B}$.
    \State \textbf{Validate.} AST-sandbox compile + field validation of $\pi_0$; on error, feed the message back and repair (up to $N_r$ attempts). If it still fails, $\text{fail}\leftarrow\text{fail}+1$; \textbf{continue}.
    \State \textbf{Search.} $(\mu{+}\lambda)$-ES from $\mu$ parents seeded by $\{\pi_0\}$, graded by the self-authored fitness $f_{\mathrm{self}}$ (frozen at generation 0); the inner scenario batch is stratified across fleet-size strata (BandedWindowSampler) and rotated every generation, so a skill cannot win by fitting one scale or one hour.
    \For{$e=1$ \textbf{to} $N_{\mathrm{gen}}$}
        \State Generate $\lambda$ children via LLM mutation, crossover, or parentless rewrite.
        \State Roll out each variant $\pi_l$ on the sampled task batch $\mathcal{T}$; grade by the fixed fitness $f_{\mathrm{self}}\bigl(\mathrm{metrics}(\pi_l)\bigr)$.
        \State Select top-$\mu$ parents.
        \State \textbf{Adaptive stop.} If $e\ge N_{\min}$ and the same leader has held for $N_{\mathrm{pat}}$ consecutive generations, stop early.
    \EndFor
    \State \textbf{Dedup.} If the champion's episode-metric signature has cosine similarity at least $\tau$ to that of an existing member of $\mathcal{B}$, $\text{fail}\leftarrow\text{fail}+1$; \textbf{continue}.
    \State \textbf{Self-check.} Audit the champion against its stated intent (see Appendix~\ref{app:prompt-skill-audit}).
    \If{verdict $=$ \textsc{match}}
        \State Freeze: add the champion to $\mathcal{B}$.
    \ElsIf{verdict $=$ \textsc{description\_wrong}}
        \State Rewrite the description card to match the measured behaviour; freeze (no re-search).
    \ElsIf{verdict $=$ \textsc{fitness\_wrong}}
        \State Re-author the fitness from the audit complaint and re-run the search (up to $N_a$ attempts); freeze on success, otherwise stamp and freeze.
    \EndIf
    \State If the audit mechanism itself fails, freeze the skill with an error stamp rather than discard it.
\EndWhile
\end{algorithmic}
\end{algorithm}

\begin{algorithm}[t!]
\caption{Phase-2 Combiner Evolution: task sampling, coverage audit, child generation, group-relative evaluation, family-elite selection, and runoff.}
\label{alg:phase2}
\begin{algorithmic}[1]
\Require Frozen skill repository $\mathcal{B}$; task distribution $\{\task_m\}$ with rewards $\mathrm{R}_m$ and dynamics $\mathcal{P}_m$; population sizes $\mu,\lambda$; crossover probability $p_{\mathrm{cross}}$; min generations $N_{\min}$; patience $N_{\mathrm{pat}}$.
\Ensure Frozen combiner $\pi_c$ that reads an unseen objective $w$ zero-shot.
\State Initialize a population of $\mu+\lambda$ proposals; sandbox-validate each; a proposal must also pass the \emph{probe-coverage} check: its code must reference every axis-specific event field (completion, dispatch wait, pickup time, detour on a new order, detour on onboard orders, solo/service time, empty move, idle wait), or it is rejected into the repair loop -- a probe set that never touches a field cannot read a reward that prices it.
\While{(generation $< N_{\min}$) \textbf{or} (no leader has repeated for $N_{\mathrm{pat}}$ consecutive generations \textbf{and} generation budget remains)}
    \State \textbf{Task sampling.} Draw a batch $\mathcal{T}=\{\task_m\}_{m=1}^{|\mathcal{T}|}$: reward families crossed with full-hour demand windows and balanced across fleet-size strata; each authored reward's coefficients are normalised so they sum to one (a global rescale that preserves the author's term ratios).
    \State \textbf{Coverage audit.} Probe each sampled objective on term-isolating events and log which metric axes the batch prices, together with the scene span (fleet bands, regimes, distinct windows); a batch whose objectives price too few axes is reported. Advisory: the audit is software-only and never blocks the round.
    \State \textbf{Child generation.} From the $\mu$ survivors produce $\lambda$ children: LLM crossover of two survivors w.p.~$p_{\mathrm{cross}}$, else single-parent mutation; one child per generation is a fresh random injection. Sandbox + field + probe-coverage validation, one repair attempt on failure.
    \For{each candidate $\pi_l$ and each task $\task_m$}
        \State $A_{l,m}\leftarrow \dfrac{\ret_l(\task_m)-\mu^{(m)}}{\sigma^{(m)}}$,
        \Statex \qquad where $\mu^{(m)}$ and $\sigma^{(m)}$ are the mean and std of the returns $\{\ret_l(\task_m)\}_l$ over the whole group evaluated on task $\task_m$.
    \EndFor
    \For{each candidate $\pi_l$}
        \State $F_l\leftarrow \frac{1}{|\mathcal{T}|}\sum_{m=1}^{|\mathcal{T}|} A_{l,m}$.
    \EndFor
    \State \textbf{Selection.} Keep the top-$\mu$ by $F_l$, with one reserved elite slot per reward family (so a hard-family specialist survives as crossover material); record the generation leader.
    \State Re-roll the surviving parents on a fresh task batch in the next generation (paired, within-round comparison).
\EndWhile
\State \textbf{Runoff.} Collect all distinct round-leaders, re-roll them together on one fresh batch, and select the final champion by a single within-comparison. Skipped if only one distinct leader exists.
\end{algorithmic}
\end{algorithm}

\begin{algorithm}[t!]
\caption{Phase-3 Repositioner Evolution: delta fitness, fairness-tripled tasks, and per-band elites.}
\label{alg:phase3}
\begin{algorithmic}[1]
\Require Frozen skill repository $\mathcal{B}$; frozen combiner $\pi_c$; reposition-off baseline; fairness strengths $\{\rho_0,\rho_{\mathrm{mid}},\rho_{\mathrm{max}}\}$; $\mu,\lambda$; min generations $N_{\min}$; patience $N_{\mathrm{pat}}$.
\Ensure Frozen reposition scorer $\pi_r$ that best-responds along the efficiency--fairness axis.
\State Initialize $\mu+\lambda$ scorer proposals; sandbox-validate each; a proposal must call $w$ in its code (a scorer that never reads the objective is rejected into the repair loop) and state its objective-response mechanism in \lstinline{objective_read_check}.
\While{(generation $< N_{\min}$) \textbf{or} (no leader has repeated for $N_{\mathrm{pat}}$ consecutive generations \textbf{and} generation budget remains)}
    \State \textbf{Task sampling.} Draw a batch $\mathcal{T}=\{\task_m\}_{m=1}^{|\mathcal{T}|}$ (same coverage audit as Algorithm~\ref{alg:phase2}) and cross it with the fairness strengths, tripling the cells $\{(\rho,\task_m)\}$.
    \State \textbf{Child generation.} Mutation / crossover / injection with sandbox validation and repair, as in Algorithm~\ref{alg:phase2}.
    \For{each candidate $\pi_l$ and each cell $(\rho,\task_m)$}
        \State $\Delta_{l}\leftarrow \ret_l^{\mathrm{repo}}(\task_m;\rho)-\ret^{\mathrm{off}}(\task_m)$;\quad $A_{l,(\rho,m)}\leftarrow \dfrac{\Delta_{l}-\mu^{(\rho,m)}}{\sigma^{(\rho,m)}}$,
        \Statex \qquad where $\ret_l^{\mathrm{repo}}(\task_m;\rho)$ and $\ret^{\mathrm{off}}(\task_m)$ are returns rolled on the same seed with repositioning on (using candidate $\pi_l$, at fairness strength $\rho$) versus off, and $\mu^{(\rho,m)}$, $\sigma^{(\rho,m)}$ are the mean and std of the deltas $\{\Delta_l\}_l$ over the whole group on this cell.
    \EndFor
    \State $F_l\leftarrow \frac{1}{3|\mathcal{T}|}\sum_{(\rho,\task_m)} A_{l,(\rho,m)}$.
    \State \textbf{Selection.} Keep the top-$\mu$ with one reserved elite per objective family AND per fairness-strength band; record the leader. Each round also reports two diagnostics per candidate: objective blindness and fairness blindness (0 = the target-region mix moves when that axis moves, 1 = it never moves).
\EndWhile
\State \textbf{Runoff} as in Algorithm~\ref{alg:phase2}.
\end{algorithmic}
\end{algorithm}

\begin{algorithm}[t!]
\caption{Runtime Dispatch: one decision step with skill blending, fairness budgets, bipartite matching, and sequential repositioning.}
\label{alg:runtime}
\begin{algorithmic}[1]
\Require Objective $w$; contexts $\phi_{\mathrm{ep}},\phi_{\step}$; skills $\mathcal{B}$; combiner $\pi_c$; repositioner $\pi_r$; fairness strength $\rho$; shared per-region state $\kappa$.
\Ensure Committed order--vehicle assignments and idle-vehicle relocations for this step.
\State \textbf{Fairness budgets.} For each vehicle $i$: $z_i\leftarrow \dfrac{E_i-\bar E}{\sigma_E+\epsilon}$; $\beta_i\leftarrow\exp(-\rho\, z_i)$.
\For{each vehicle $i$}
    \State $\alpha_i\leftarrow \pi_c(\obs_i,\phi_{\mathrm{ep}},\phi_{\step},w)$; keep the top-$b$ positive scores as $\mathcal{K}_i$; softmax-normalize to $\{\tilde\alpha_{i,k}\}_{k\in\mathcal{K}_i}$ (Eq.~\eqref{eq:blend-weights}).
    \For{each feasible candidate order $o_j$, $j\in\mathcal{J}_i$}
        \State z-normalize each retained skill over $i$'s candidates: $\hat s_k\leftarrow \dfrac{s_k(\obs_i,o_j,\phi_{\mathrm{ep}},\phi_{\step})-\mu_k}{\sigma_k+\epsilon}$.
        \State $a_{i,j}\leftarrow \beta_i\sum_{k\in\mathcal{K}_i} \tilde\alpha_{i,k}\,\hat s_k$.
    \EndFor
\EndFor
\State \textbf{Matching.} Solve the bipartite matching program of Eq.~\eqref{eq:matching} with scores $a_{i,j}$ (including the no-op option $\varnothing$), committing each pending order to at most one vehicle and each vehicle to at most one new order.
\State \textbf{Sequential repositioning.}
\For{each idle, unmatched vehicle $i$ in a random order}
    \State Score candidate regions via $\{\nu_g\}_{g\in\mathcal{G}_i}\leftarrow\pi_r(\obs_i,\phi_{\mathrm{ep}},\phi_{\step},\kappa,w,\beta_i)$; select $g^*=\arg\max_{g\in\mathcal{G}_i}\nu_g$ subject to the minimum-relocation-benefit stay rules.
    \State Update $\kappa$: decrement $\kappa_{g^*}$ before the next vehicle scores, so later cars see the demand the earlier ones already claimed.
\EndFor
\end{algorithmic}
\end{algorithm}

\section{Experiment Details} \label{app:exp}
\subsection{Training Configurations}
\label{sec:training-config}

Table~\ref{tab:config} lists the hyperparameters of the three training phases. All phases share the $(\mu{+}\lambda)$-ES with group-relative (GDPO-style) fitness. The experiments are conducted on a workstation running Windows 11, equipped with an Intel(R) Core(TM) i7-14700KF processor without GPU using.

\begin{table}[htbp]
\centering
\caption{Training hyperparameters. ``Shared'' refers to the common $(\mu{+}\lambda)$-ES setup; the per-phase rows list the settings of that phase's search.}
\label{tab:config}
\smallskip
\adjustbox{max width=\textwidth}{%
\begin{tabular}{@{} l l l @{}}
\toprule
\textbf{Phase} & \textbf{Hyperparameter} & \textbf{Value} \\
\midrule
\multirow{6}{*}{Shared}
 & parent / child population $(\mu{+}\lambda)$ & $(4{+}4)$ \\
 & LLM crossover probability $p_{\mathrm{cross}}$ & $0.35$ \\
 & parentless fresh injection & 1 child per generation \\
 & sandbox repair attempts & $\le 3$ (cooling temperature) \\
 & rollout workers & 8 \\
 & generation temperature &  $0.9$ \\
\midrule
\multirow{5}{*}{Phase 1 (skills)}
 & repository capacity $K$ & $10$  \\
 & generations per skill & $5$ \\
 & scenarios per generation & $6$  \\
 & behavioral deduplication threshold & cosine similarity $\tau = 0.98$ \\
 & proposed skills & 20 proposals \\
\midrule
\multirow{4}{*}{Phase 2-3 (combiner and repositioner)}
 & generations & $8\sim20$ (adaptive stop: early stop after 3 stable rounds) \\
 & tasks per generation & $18$ (scenario, objective) \\
 & objective mixing & $0.5$ structural-family fraction \\
 & elite & 1 reserved per reward family \\
\bottomrule
\end{tabular}%
}
\end{table}

\subsection{Comparative Baselines}
\label{sec:baselines}

We compare against three families of baselines: model-based heuristics, MARL agents, and an LLM-based dispatch method. Each is described below. All MARL agents are trained with the ride-sharing gym of \citet{zhao2026ridegym} on the same anchor objective and scenario as our method.

\textbf{Nearest distance} \citep{kalyanasundaram1993online} assigns each pending order to the closest idle vehicle. Candidate vehicles are retrieved by spatial hashing for roughly constant-time lookup, and ties are broken by the order's arrival time. 

\textbf{Kuhn-Munkres (KM)} \citep{kuhn1955hungarian,simonetto2019real} solves a bipartite assignment over all vehicle-order pairs at each step, minimizing the total Euclidean pickup distance. Capacity-infeasible pairs are excluded by assigning them cost $+\infty$; the resulting assignment is globally minimal for the current step.

\textbf{Gale-Shapley (GS)} \citep{gale1962college,yue2024end} computes a stable one-to-one matching via deferred acceptance, in which orders propose to vehicles in order of pickup distance and each vehicle keeps the best offer it has received so far, iterating until no mutually preferable swap remains. 

\textbf{REDA} \citep{holder2025multi} is an independent IDDQN-style baseline. Each vehicle is an independent learner whose Q-network estimates the value of assigning a candidate order, and the fleet-level assignment is recovered by bipartite matching over the per-vehicle Q-values. 

\textbf{BMG-Q} \citep{hu2025bmg} extends IDDQN with a localized bipartite-match graph attention module. The Q-value of a vehicle-order pair is computed by a graph attention encoder that aggregates the states of a vehicle's nearby agents and orders, so each learner conditions on a local neighborhood while still acting independently. 

\textbf{MFRL} \citep{li2019efficient} replaces the attention encoder with a mean-field approximation. Each vehicle's Q-value depends on the aggregate behavior of its neighbors, summarized by a mean-field action computed from the average order information in the neighborhood. This captures local interaction implicitly without resorting to the full joint action space.

\textbf{\citet{zhang2026hierarchical}} generates dispatch heuristics offline by combining LLM generation with evolutionary search. We evaluate both reported variants: an open-loop version that runs the evolved heuristic without any intermediate LLM call, and a close-loop version that re-invokes the LLM during deployment. This is the closest prior work to ours in spirit (LLM-aided automatic design), but it targets ride-hailing without vehicle sharing.

\subsection{Evaluation Metrics}
\label{sec:metrics}

All tables and figures in the paper report the same metric set, computed by the benchmark recorder from the raw episode log. Unless stated otherwise, per-order quantities are averaged over the orders of one episode and then reported as mean $\pm$ standard deviation across evaluation scenarios (windows, fleets, or axis levels, depending on the experiment).

\begin{itemize}[left=0pt]
\item \textbf{Reward.} The cumulative value of the episode's injected reward function, summed over every vehicle and every decision step. All methods in a comparison are evaluated using the same injected reward, so values are directly comparable within a table. Across different objectives, however, the reward scale follows the objective's own price list and is not comparable between different reward functions.

\item \textbf{Service Rate.} The fraction of orders that are ever assigned to a vehicle, defined as $\mathrm{confirmed} / \mathrm{total}$. This metric measures dispatch coverage: an order is counted once it is committed to a vehicle, regardless of whether the ride is completed within the episode horizon.

\item \textbf{Completion Rate.} The fraction of orders that are actually delivered (dropped off) within the horizon, defined as $\mathrm{completed} / \mathrm{total}$. This value is always less than or equal to the service rate; the gap corresponds to orders that remain en route when the episode ends.

\item \textbf{Wait Time (min).} Per-order waiting time from the moment the order is placed to the moment the rider boards ($t_{\mathrm{pickup}} - t_{\mathrm{request}}$), averaged over all orders. This includes the full out-of-vehicle wait, covering both the time until the platform commits a vehicle and the vehicle's travel to the pickup point. Canceled orders contribute their time until cancellation.

\item \textbf{Ride Time (min).} Per-order in-vehicle time from boarding to drop-off ($t_{\mathrm{dropoff}} - t_{\mathrm{pickup}}$), averaged over completed orders.

\item \textbf{Detour Time (min).} Per-order pooling detour, defined as the realized ride time minus the direct origin-to-destination travel time on the road network, clamped at zero and averaged over completed orders. This captures the extra in-vehicle time a rider experiences due to serving other passengers along the route; an unpooled ride has a detour time of $0$.

\item \textbf{Utilization.} Mean vehicle utilization, defined as the fraction of decision steps during which a vehicle is busy (serving or traveling to serve at least one order), averaged over the fleet. Idle waiting and empty repositioning are both counted as non-busy.

\end{itemize}

The fairness experiments (Table~\ref{tab:fairness}) additionally report per-vehicle dispersion statistics. Each is the standard deviation of a per-vehicle quantity across the fleet; lower values indicate greater equity:
\begin{itemize}[left=0pt]
\item \textbf{Orders Std:} standard deviation of the number of orders assigned to each vehicle.
\item \textbf{Served Std:} standard deviation of the number of passengers each vehicle actually served.
\item \textbf{Dist. Std;} standard deviation of each vehicle's total travel distance (in meters).
\item \textbf{W-Dist Std:} standard deviation of each vehicle's passenger-weighted travel distance, defined as $d_v \cdot p_v / \sum_u p_u$, where $d_v$ is vehicle $v$'s travel distance and $p_v$ is the number of passengers it served. This metric is motivated by the fact that in ride-sharing, the fare is proportional to the number of passengers.
\end{itemize}

\subsection{Objective Response}
\label{sec:objective}

In this section, we illustrate how our method adapts the policy to different objectives without retraining—a capability not supported by previous approaches, whose policies remain fixed regardless of the objective. Each evaluation point is obtained by rolling out five held-out test windows, with results reported as mean $\pm$ standard deviation across windows, using the same metric set as in Table~\ref{tab:generalization}. For the comparison table, we additionally run an identical stack without providing the objective to the combiner (i.e., the combiner cannot read the objective), serving as a blind control under the same injected reward. Since both arms are evaluated under the same reward function, their performance is directly comparable, and the observed gap precisely measures the benefit contributed by reading the objective.

\paragraph{Impact of Term Weight in Reward Function}

\begin{figure}[t!]
    \centering
        \begin{subfigure}{\textwidth}
        \centering
        \includegraphics[width=\textwidth]{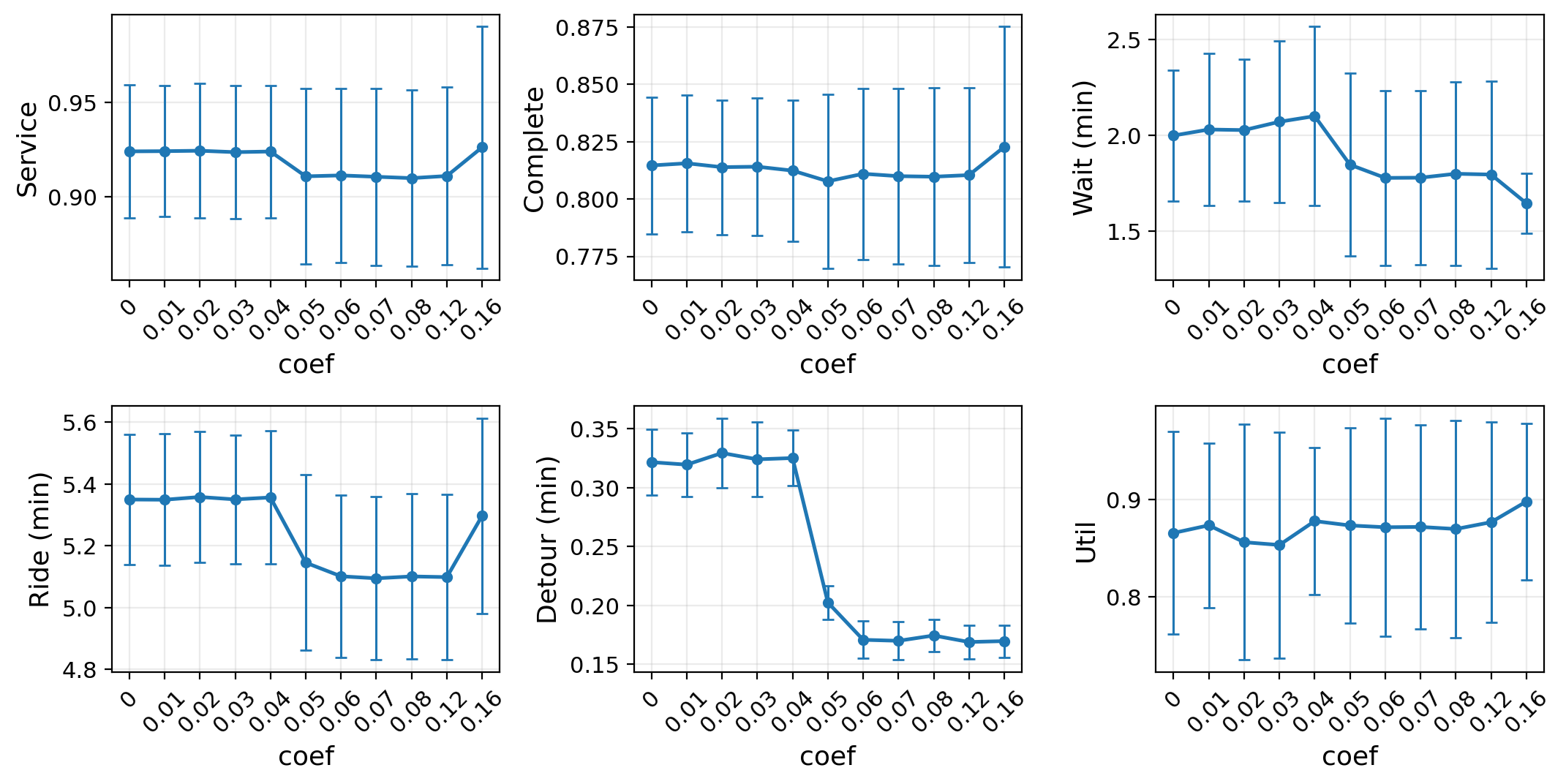}
        \caption{Detour Coefficient Sweep}
        \label{fig:detour_sweep}
    \end{subfigure}
        \begin{subfigure}{\textwidth}
        \centering
        \includegraphics[width=\textwidth]{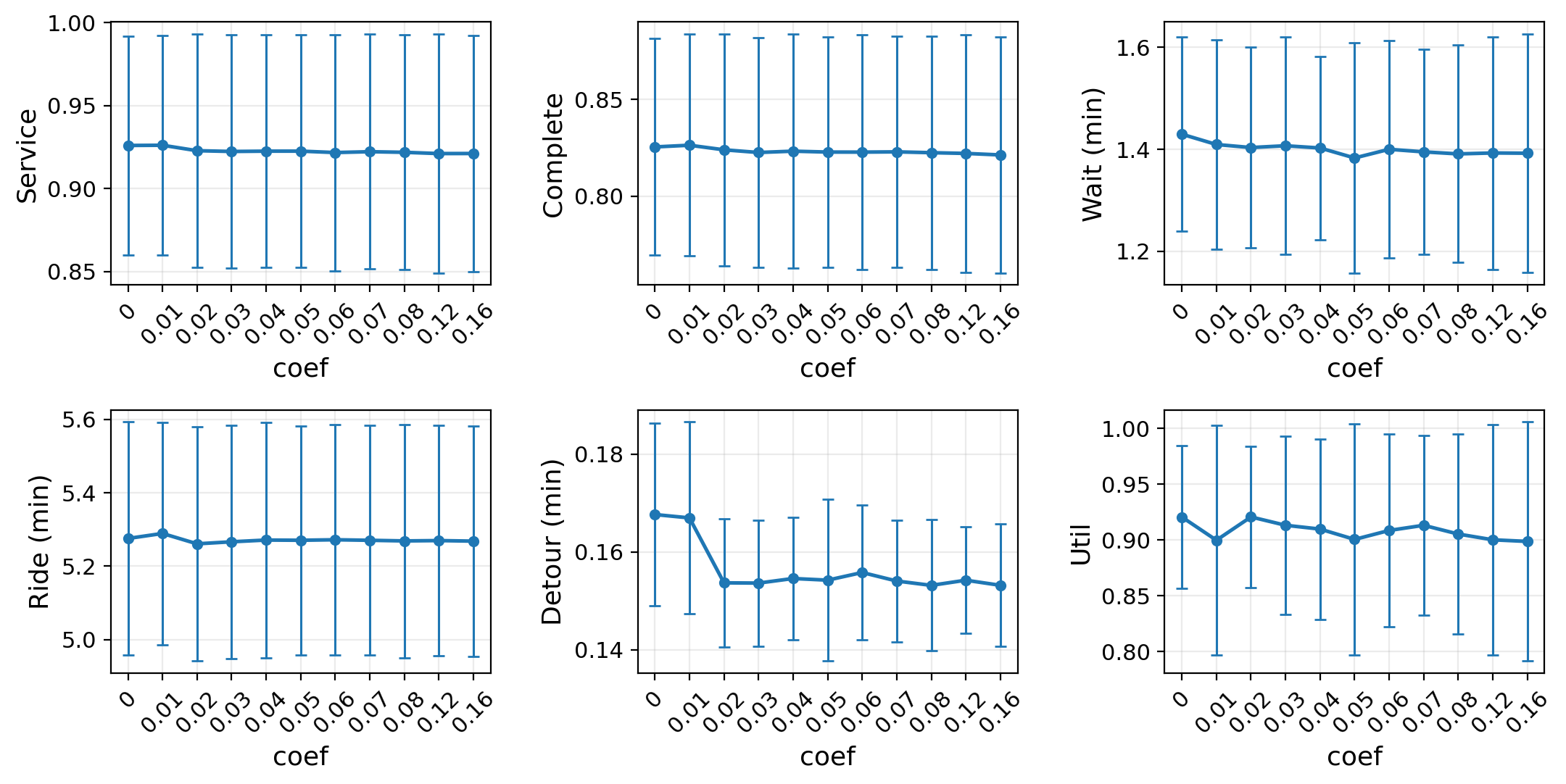}
        \caption{Pickup-Time Coefficient Sweep}
        \label{fig:pickup_sweep}
    \end{subfigure}
    \begin{subfigure}{\textwidth}
        \centering
        \includegraphics[width=\textwidth]{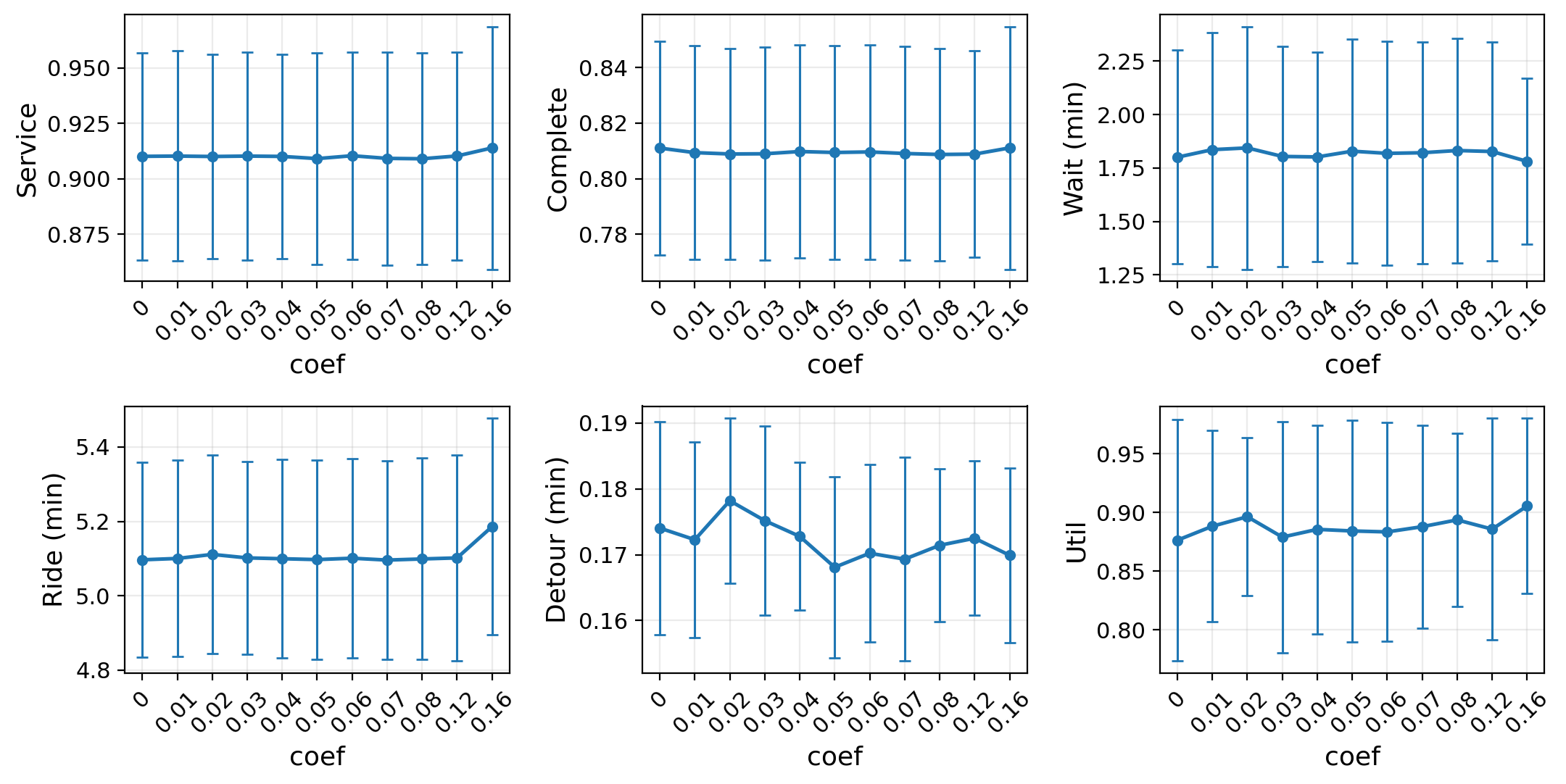}
        \caption{Service-Time Sweep}
        \label{fig:service_sweep}
    \end{subfigure}
    \caption{The six evaluation metrics under varying detour, pickup-time, and service-time coefficients.}
    \label{fig:combined_sweeps}
\end{figure}

In this part, we examine the fine-grained policy changes induced by sweeping individual coefficients in the reward function. Figure~\ref{fig:combined_sweeps} presents the corresponding metric variations when the coefficients of detour time, pickup time (i.e., wait time), and total ride time are swept separately.

The effects in Fig.~\ref{fig:detour_sweep} and Fig.~\ref{fig:pickup_sweep} are clear and direct: as the coefficient increases, both detour time and wait time decrease, reflecting greater emphasis placed on these terms by the platform. In contrast, the impact in Fig.~\ref{fig:service_sweep} is less straightforward but still interpretable. As the coefficient of service time (ride time) increases, the policy tends to reduce detour and pickup times in order to improve overall ride time, which explains the observed decrease in both metrics. Under this setting, the platform also benefits from increased idle labor capacity, enabling it to serve more orders—particularly those with longer travel distances. This is supported by the observation that while detour time decreases, ride time increases, suggesting that the platform allocates more resources to fulfilling longer trips. Consequently, the average ride time does not change significantly, while the service rate exhibits a slight increase.

\paragraph{Impact of Given Objective}

\begin{table}[htbp]
\centering
\caption{For each objective, the identical stack is rolled both with the objective (i.e., reads $w$) and without it (i.e., blind, $w = \text{None}$), on the same five test windows under the same injected reward. The only difference is whether the combiner reads the objective at inference time. Top two columns correspond to given reward-function objectives; bottom two correspond to natural-language (NL) objectives authored by the model from a brief. Bold values highlight metrics that align with the given objective.}
\label{tab:w_delta}
\smallskip
\adjustbox{max width=\textwidth}{%
\begin{tabular}{@{} l r r r r r r r r @{}}
\toprule
 & \multicolumn{2}{c}{\textbf{pickup-time}} & \multicolumn{2}{c}{\textbf{detour-time}} & \multicolumn{2}{c}{\textbf{NL long-trip}} & \multicolumn{2}{c}{\textbf{NL detour-time}} \\
\textbf{Metric} & \textbf{blind} & \textbf{reads $w$} & \textbf{blind} & \textbf{reads $w$} & \textbf{blind} & \textbf{reads $w$} & \textbf{blind} & \textbf{reads $w$} \\
\midrule
Service  & 0.93$\pm$0.04 & 0.92$\pm$0.07 & 0.93$\pm$0.04 & 0.91$\pm$0.05 & 0.93$\pm$0.04 & 0.89$\pm$0.10 & 0.93$\pm$0.04 & 0.90$\pm$0.05 \\
Complete & 0.82$\pm$0.03 & 0.82$\pm$0.06 & 0.82$\pm$0.03 & 0.81$\pm$0.04 & 0.82$\pm$0.03 & 0.72$\pm$0.07 & 0.82$\pm$0.03 & 0.80$\pm$0.04 \\
Wait     & \underline{1.66}$\pm$\underline{0.13} & \textbf{1.39}$\pm$\textbf{0.21} & 1.66$\pm$0.13 & 1.80$\pm$0.48 & 1.66$\pm$0.13 & 3.37$\pm$0.56 & {1.66}$\pm${0.13} & {2.07}$\pm${0.76} \\
Ride     & \underline{5.33}$\pm$\underline{0.20} & 5.27$\pm$0.32 & 5.33$\pm$0.20 & 5.10$\pm$0.27 & 5.33$\pm$0.20 & \textbf{7.07}$\pm$\textbf{0.44} & 5.33$\pm$0.20 & 5.10$\pm$0.24 \\
Detour   & 0.30$\pm$0.04 & 0.15$\pm$0.01 & \underline{0.30}$\pm$\underline{0.04} & \textbf{0.17}$\pm$\textbf{0.01} & 0.30$\pm$0.04 & 1.75$\pm$0.19 & \underline{0.30}$\pm$\underline{0.04} & \textbf{0.16}$\pm$\textbf{0.01} \\
Util     & 0.84$\pm$0.14 & 0.91$\pm$0.09 & 0.84$\pm$0.14 & 0.87$\pm$0.11 & 0.84$\pm$0.14 & 0.89$\pm$0.09 & 0.84$\pm$0.14 & 0.88$\pm$0.09 \\
\bottomrule
\end{tabular}}
\end{table}

In this part, we explore how the policy changes in response to different given objectives, whether specified via a direct mathematical reward function or through a natural-language (NL) description. The results are presented in Table~\ref{tab:w_delta}, where we observe that the policy adapts accordingly for each objective.

For the mathematical reward functions (minimizing pickup time and minimizing detour time), we observe that the corresponding metrics decrease as expected. For the NL-based objective, we first instruct the policy to prioritize serving long-trip orders, which typically yield higher revenue. We observe that ride time increases by 1.74 minutes while detour time increases by 1.45 minutes, indicating that the remaining 0.19-minute difference stems from policy adjustments. However, this strategy inevitably increases detour time, as the majority of orders along the route are short-distance ones. Finally, we test the policy with an NL objective to minimize detour time. The results show that this NL-driven objective achieves performance comparable to that of the mathematically specified counterpart, suggesting that the LLM correctly interprets the natural-language objective on its own.

\subsection{Fairness}
\label{sec:fairness}

We study the efficiency-fairness tradeoff by varying the fairness strength $\rho$ from 0 (disabled) to 1.0. Table~\ref{tab:fairness} reports the performance metrics and per-vehicle fairness standard deviations (std, lower is more equitable) for all baselines, MARL agents, and RideSkill variants. Each cell is evaluated across five in-domain-adjacent scenarios (fleet sizes 800, 1000, and 1200; hours 17, 18, and 19). We observe that our method achieves overall better fairness compared to others, which may be attributed to the inclusion of fairness-related skills within the skill repository itself. We then note that our proposed heuristic fairness budget mechanism is particularly effective in the policy without repositioning: increasing $\rho$ helps reduce the metric deviation among vehicles without significantly impacting overall performance. However, this mechanism is less effective when the repositioner is active, as the repositioner already directly improves fairness metrics even without the fairness budget mechanism. This is because the repositioner balances supply and demand by relocating idle vehicles—typically those with lower income—to high-demand regions. Nevertheless, we believe this mechanism remains meaningful, as some platforms may not have full control over vehicle movements, and repositioning may not always be supported.

\begin{table}[htbp]
\centering
\caption{The efficiency-fairness tradeoff, mean$\pm$std across five scenarios.  Left: performance metrics. Right: per-vehicle fairness std, where lower is more equitable.}
\label{tab:fairness}
\smallskip
\adjustbox{max width=\textwidth}{%
\begin{tabular}{@{} l r r r r r r r r @{}}
\toprule
 & \multicolumn{3}{c}{\textbf{Performance}} & & \multicolumn{4}{c}{\textbf{Fairness std}} \\
\textbf{Configuration} & \textbf{Reward} & \textbf{Service} & \textbf{Complete} & & \textbf{Orders} & \textbf{Served} & \textbf{Dist.} & \textbf{W-Dist.} \\
\midrule
nearest distance & 6{,}779$\pm$867 & 0.792$\pm$0.068 & 0.652$\pm$0.060 & & 2.45$\pm$0.17 & 2.31$\pm$0.14 & 3{,}916$\pm$1{,}269 & 2{,}964$\pm$654 \\
KM & 6{,}091$\pm$659 & 0.749$\pm$0.054 & 0.607$\pm$0.047 & & 3.40$\pm$0.39 & 2.96$\pm$0.28 & 9{,}787$\pm$1{,}933 & 4{,}649$\pm$237 \\
GS & 5{,}330$\pm$749 & 0.644$\pm$0.074 & 0.512$\pm$0.064 & & 3.21$\pm$0.27 & 2.67$\pm$0.21 & 10{,}787$\pm$1{,}110 & 5{,}055$\pm$342 \\
REDA & 7{,}257$\pm$1{,}095 & 0.809$\pm$0.078 & 0.662$\pm$0.071 & & 2.16$\pm$0.06 & 2.00$\pm$0.04 & 3{,}333$\pm$705 & 2{,}601$\pm$425 \\
MFRL & 7{,}365$\pm$1{,}157 & 0.823$\pm$0.086 & 0.678$\pm$0.079 & & 2.06$\pm$0.08 & 1.90$\pm$0.06 & 1{,}657$\pm$286 & 1{,}596$\pm$279 \\
BMG-Q & 7{,}274$\pm$1{,}002 & 0.807$\pm$0.073 & 0.662$\pm$0.066 & & 2.15$\pm$0.05 & 1.99$\pm$0.05 & 2{,}500$\pm$417 & 2{,}281$\pm$350 \\
\midrule
\multicolumn{9}{l}{\emph{RideSkill (w/o reps) + fairness budget}} \\
$\rho{=}0$ & 9{,}328$\pm$733 & 0.850$\pm$0.059 & 0.762$\pm$0.049 & & 1.84$\pm$0.10 & 1.83$\pm$0.09 & 2{,}114$\pm$644 & 1{,}904$\pm$500 \\
$\rho{=}0.25$ & 9{,}361$\pm$723 & 0.854$\pm$0.055 & 0.763$\pm$0.048 & & 1.64$\pm$0.06 & 1.64$\pm$0.05 & 1{,}904$\pm$369 & 1{,}856$\pm$346 \\
$\rho{=}0.50$ & 9{,}380$\pm$704 & 0.857$\pm$0.050 & 0.766$\pm$0.044 & & 1.67$\pm$0.02 & 1.67$\pm$0.00 & 1{,}815$\pm$310 & 1{,}799$\pm$298 \\
$\rho{=}0.75$ & 9{,}368$\pm$690 & 0.859$\pm$0.048 & 0.765$\pm$0.042 & & 1.69$\pm$0.04 & 1.70$\pm$0.03 & 1{,}766$\pm$262 & 1{,}761$\pm$255 \\
$\rho{=}1.0$ & 9{,}363$\pm$677 & 0.859$\pm$0.046 & 0.765$\pm$0.041 & & 1.75$\pm$0.04 & 1.76$\pm$0.03 & 1{,}710$\pm$239 & 1{,}715$\pm$238 \\
\midrule
\multicolumn{9}{l}{\emph{RideSkill (full stack)}} \\
$\rho{=}0$ & 9{,}470$\pm$776 & 0.859$\pm$0.061 & 0.770$\pm$0.051 & & 1.66$\pm$0.06 & 1.67$\pm$0.05 & 957$\pm$103 & 949$\pm$96 \\
$\rho{=}0.25$ & 9{,}412$\pm$769 & 0.857$\pm$0.060 & 0.767$\pm$0.050 & & 1.63$\pm$0.05 & 1.64$\pm$0.03 & 1{,}057$\pm$109 & 1{,}063$\pm$107 \\
$\rho{=}0.50$ & 9{,}371$\pm$754 & 0.854$\pm$0.058 & 0.764$\pm$0.048 & & 1.72$\pm$0.05 & 1.74$\pm$0.04 & 1{,}100$\pm$111 & 1{,}113$\pm$109 \\
$\rho{=}0.75$ & 9{,}366$\pm$746 & 0.853$\pm$0.057 & 0.764$\pm$0.047 & & 1.73$\pm$0.08 & 1.74$\pm$0.06 & 1{,}108$\pm$106 & 1{,}119$\pm$107 \\
$\rho{=}1.0$ & 9{,}306$\pm$752 & 0.850$\pm$0.057 & 0.763$\pm$0.047 & & 1.85$\pm$0.09 & 1.86$\pm$0.06 & 1{,}119$\pm$129 & 1{,}134$\pm$133 \\
\bottomrule
\end{tabular}}
\end{table}

\subsection{Evaluation on Open-Source LLM}

To evaluate whether RideSkill's training pipeline generalizes beyond proprietary models, we applied the full three-phase pipeline using GLM-5.1 \citep{zeng2026glm} as the LLM operator, replacing the Claude Opus 4.8 model used in the main experiments. Table~\ref{tab:glm-generalization} compares the two models on the five-axis generalization sweep from Table~\ref{tab:generalization}, for the same three stacks: single-skill average, \emph{RideSkill (w/o repos)}, and \emph{RideSkill}. The results show that Claude Opus 4.8 outperforms GLM-5.1 across all scenarios, suggesting that the performance of our pipeline is influenced by the capacity of the underlying LLM. Nevertheless, we observe that GLM-5.1 still surpasses all benchmarks in out-of-domain settings, demonstrating the robustness of our framework design, which enables the model to successfully adapt to diverse scenarios.

\begin{table}[htbp]
\centering
\caption{Generalization across five scenario axes for the Claude Opus 4.8 and GLM-5.1 models. Each cell is the mean$\pm$std of the raw metric across all levels of the given axis.}
\label{tab:glm-generalization}
\smallskip
\adjustbox{max width=\textwidth}{%
\begin{tabular}{@{} l l l r r r r r r r @{}}
\toprule
\textbf{Axis} & \textbf{Model} & \textbf{Method} & \textbf{Reward} & \textbf{Service} & \textbf{Complete} & \textbf{Wait} & \textbf{Ride} & \textbf{Detour} & \textbf{Util} \\
\midrule
\multirow{6}{*}{\rotatebox{90}{\small\textbf{Hour}}} & \multirow{3}{*}{Claude Opus 4.8 } & single-skill & 4{,}876$\pm$970 & 0.72$\pm$0.05 & 0.61$\pm$0.04 & 2.36$\pm$0.08 & 5.91$\pm$0.20 & 1.12$\pm$0.13 & 0.61$\pm$0.11 \\
&  & RideSkill (w/o reps) & 7{,}221$\pm$1{,}360 & 0.88$\pm$0.07 & 0.79$\pm$0.06 & 1.14$\pm$0.07 & 5.05$\pm$0.27 & 0.17$\pm$0.02 & 0.71$\pm$0.12 \\
&  & RideSkill & 7{,}654$\pm$1{,}497 & 0.92$\pm$0.06 & 0.83$\pm$0.05 & 1.12$\pm$0.15 & 5.24$\pm$0.24 & 0.16$\pm$0.02 & 0.90$\pm$0.10 \\
& \multirow{3}{*}{GLM-5.1 } & single-skill & 5{,}755$\pm$1{,}113 & 0.86$\pm$0.07 & 0.72$\pm$0.05 & 2.96$\pm$0.22 & 6.15$\pm$0.15 & 1.14$\pm$0.12 & 0.73$\pm$0.12 \\
&  & RideSkill (w/o reps) & 6{,}045$\pm$1{,}206 & 0.89$\pm$0.06 & 0.76$\pm$0.05 & 1.46$\pm$0.07 & 6.06$\pm$0.29 & 1.37$\pm$0.05 & 0.64$\pm$0.12 \\
&  & RideSkill & 6{,}406$\pm$1{,}377 & 0.93$\pm$0.03 & 0.79$\pm$0.03 & 1.39$\pm$0.13 & 6.27$\pm$0.21 & 1.40$\pm$0.05 & 0.88$\pm$0.10 \\
\midrule
\multirow{6}{*}{\rotatebox{90}{\small\textbf{Fleet}}} & \multirow{3}{*}{Claude Opus 4.8 } & single-skill & 4{,}835$\pm$1{,}862 & 0.49$\pm$0.18 & 0.42$\pm$0.16 & 2.06$\pm$0.06 & 6.16$\pm$0.25 & 1.36$\pm$0.16 & 0.73$\pm$0.07 \\
&  & RideSkill (w/o reps) & 7{,}517$\pm$2{,}638 & 0.64$\pm$0.22 & 0.58$\pm$0.19 & 1.16$\pm$0.08 & 4.14$\pm$0.69 & 0.19$\pm$0.01 & 0.85$\pm$0.08 \\
&  & RideSkill & 7{,}907$\pm$2{,}945 & 0.67$\pm$0.24 & 0.61$\pm$0.21 & 1.22$\pm$0.10 & 4.24$\pm$0.80 & 0.19$\pm$0.02 & 0.95$\pm$0.02 \\
& \multirow{3}{*}{GLM-5.1 } & single-skill & 5{,}723$\pm$2{,}376 & 0.59$\pm$0.23 & 0.50$\pm$0.20 & 2.74$\pm$0.12 & 6.26$\pm$0.09 & 1.41$\pm$0.14 & 0.85$\pm$0.06 \\
&  & RideSkill (w/o reps) & 6{,}496$\pm$2{,}085 & 0.65$\pm$0.22 & 0.57$\pm$0.18 & 1.43$\pm$0.04 & 5.01$\pm$0.84 & 1.21$\pm$0.15 & 0.81$\pm$0.10 \\
&  & RideSkill & 6{,}872$\pm$2{,}320 & 0.69$\pm$0.24 & 0.61$\pm$0.20 & 1.46$\pm$0.11 & 5.16$\pm$0.99 & 1.23$\pm$0.17 & 0.95$\pm$0.02 \\
\midrule
\multirow{6}{*}{\rotatebox{90}{\small\textbf{Speed}}} & \multirow{3}{*}{Claude Opus 4.8 } & single-skill & 5{,}780$\pm$1{,}126 & 0.59$\pm$0.09 & 0.50$\pm$0.11 & 2.24$\pm$0.71 & 6.39$\pm$1.68 & 1.28$\pm$0.18 & 0.70$\pm$0.03 \\
&  & RideSkill (w/o reps) & 8{,}904$\pm$1{,}151 & 0.75$\pm$0.11 & 0.68$\pm$0.12 & 1.19$\pm$0.31 & 4.81$\pm$1.13 & 0.20$\pm$0.04 & 0.82$\pm$0.04 \\
&  & RideSkill & 9{,}457$\pm$1{,}371 & 0.79$\pm$0.13 & 0.72$\pm$0.13 & 1.32$\pm$0.38 & 4.96$\pm$1.07 & 0.21$\pm$0.05 & 0.95$\pm$0.00 \\
& \multirow{3}{*}{GLM-5.1 } & single-skill & 6{,}913$\pm$1{,}468 & 0.71$\pm$0.11 & 0.60$\pm$0.13 & 2.93$\pm$0.88 & 6.64$\pm$1.72 & 1.33$\pm$0.22 & 0.83$\pm$0.04 \\
&  & RideSkill (w/o reps) & 7{,}543$\pm$1{,}268 & 0.76$\pm$0.11 & 0.66$\pm$0.12 & 1.53$\pm$0.42 & 5.75$\pm$1.06 & 1.32$\pm$0.10 & 0.76$\pm$0.03 \\
&  & RideSkill & 8{,}090$\pm$1{,}366 & 0.83$\pm$0.11 & 0.72$\pm$0.12 & 1.62$\pm$0.54 & 6.07$\pm$1.05 & 1.39$\pm$0.09 & 0.95$\pm$0.01 \\
\midrule
\multirow{6}{*}{\rotatebox{90}{\small\textbf{Cap$\leq$4}}} & \multirow{3}{*}{Claude Opus 4.8 } & single-skill & 4{,}087$\pm$1{,}531 & 0.41$\pm$0.16 & 0.36$\pm$0.14 & 1.79$\pm$0.18 & 5.60$\pm$0.36 & 0.81$\pm$0.43 & 0.51$\pm$0.18 \\
&  & RideSkill (w/o reps) & 6{,}142$\pm$2{,}422 & 0.53$\pm$0.20 & 0.48$\pm$0.18 & 1.16$\pm$0.06 & 5.00$\pm$0.32 & 0.18$\pm$0.02 & 0.61$\pm$0.19 \\
&  & RideSkill & 6{,}538$\pm$2{,}667 & 0.56$\pm$0.22 & 0.51$\pm$0.20 & 1.22$\pm$0.11 & 5.17$\pm$0.27 & 0.17$\pm$0.02 & 0.93$\pm$0.02 \\
& \multirow{3}{*}{GLM-5.1 } & single-skill & 5{,}053$\pm$1{,}800 & 0.51$\pm$0.19 & 0.44$\pm$0.16 & 2.30$\pm$0.28 & 5.81$\pm$0.47 & 0.86$\pm$0.45 & 0.63$\pm$0.20 \\
&  & RideSkill (w/o reps) & 5{,}316$\pm$1{,}970 & 0.53$\pm$0.21 & 0.47$\pm$0.18 & 1.32$\pm$0.09 & 5.54$\pm$0.32 & 0.94$\pm$0.47 & 0.56$\pm$0.19 \\
&  & RideSkill & 5{,}648$\pm$2{,}217 & 0.57$\pm$0.24 & 0.50$\pm$0.20 & 1.36$\pm$0.11 & 5.72$\pm$0.39 & 0.95$\pm$0.48 & 0.93$\pm$0.02 \\
\midrule
\multirow{6}{*}{\rotatebox{90}{\small\textbf{Cap$>$4}}} & \multirow{3}{*}{Claude Opus 4.8 } & single-skill & 5{,}413$\pm$301 & 0.65$\pm$0.02 & 0.53$\pm$0.00 & 2.29$\pm$0.11 & 6.54$\pm$0.22 & 1.97$\pm$0.27 & 0.69$\pm$0.01 \\
&  & RideSkill (w/o reps) & 9{,}144$\pm$303 & 0.77$\pm$0.00 & 0.70$\pm$0.00 & 1.10$\pm$0.00 & 4.58$\pm$0.01 & 0.23$\pm$0.00 & 0.82$\pm$0.00 \\
&  & RideSkill & 9{,}709$\pm$301 & 0.82$\pm$0.00 & 0.74$\pm$0.00 & 1.23$\pm$0.00 & 4.74$\pm$0.01 & 0.22$\pm$0.00 & 0.96$\pm$0.00 \\
& \multirow{3}{*}{GLM-5.1 } & single-skill & 6{,}768$\pm$276 & 0.80$\pm$0.03 & 0.66$\pm$0.01 & 2.96$\pm$0.12 & 6.92$\pm$0.26 & 2.07$\pm$0.31 & 0.83$\pm$0.01 \\
&  & RideSkill (w/o reps) & 7{,}192$\pm$189 & 0.84$\pm$0.03 & 0.70$\pm$0.02 & 1.43$\pm$0.08 & 6.37$\pm$0.35 & 2.07$\pm$0.27 & 0.76$\pm$0.01 \\
&  & RideSkill & 7{,}712$\pm$199 & 0.91$\pm$0.02 & 0.76$\pm$0.00 & 1.43$\pm$0.13 & 6.69$\pm$0.25 & 2.13$\pm$0.22 & 0.95$\pm$0.00 \\
\bottomrule
\end{tabular}%
}
\end{table}

\subsection{Examples}
\label{app:examples}

In this section, we present example outputs from each of the three training phases. For each artifact, we show both the generated code and the natural-language card that the LLM authored alongside it—including its stated objective, the mechanism or strategy it implements, and a description of the expected behaviour.


\subsubsection{Skill: \texttt{near\_short\_efficiency}}

\textbf{Objective (LLM):} Maximise the share of demand served by grabbing any
feasible short-detour trip, pushing the service rate and assigned volume up
rather than minimising per-passenger waiting and in-car time.

\textbf{Mechanism (LLM):} hard completability gate then rank by marginal
productive-time efficiency, with an OD-region re-demand prior tiebreak and a
region-aware option-value noop.

\textbf{Description (LLM):} A hard completability gate admits trips below a
modest solo-time budget, rejecting only the longest commitments. Survivors are
ranked by a transit-efficiency term that collapses to near-unity on almost any
viable ride and strongly favours accepting whatever is feasible, with a detour
term that drives detour per order to $\sim$0 (0.00--0.21) so distance per
delivery is small. The demand-pressure gate widens only slightly, so under real
pressure almost all flood survives; a low noop floor makes idle cars take the
first feasible order. The result is high service rate (0.361--0.922) and high
assigned volume (3840--9301) with moderate mean service time (5.68--8.17 min),
i.e. broad, busy coverage rather than few very fast trips.

\begin{lstlisting}[style=pybox, caption={Frozen skill \texttt{near\_short\_efficiency} (Phase 1, gen~3).}]
def score(driver_obs, order, phi_ep, phi_step):
    s = driver_obs["self"]
    free = s["capacity"] - s["committed_passengers"]
    if order["num_passengers"] > free:
        return -1e9
    dist = phi_ep.dist
    scale = float(phi_step.mean_solo_time)
    if scale is None or scale <= 0:
        scale = float(phi_ep.scale) or 1.0
    pickup = dist(s["location"], order["origin"]) / scale
    ride = dist(order["origin"], order["destination"]) / scale
    commit = pickup + ride

    # Hard completability gate in live solo-time units
    dp = phi_step.demand_pressure
    window = 2.0 + 2.0 * (1.0 - 1.0 / (1.0 + dp))
    if commit > window:
        return -1e9

    held = onboard = 0.0
    for d in s["assigned_order_details"]:
        if d.get("onboard", False): onboard += 1.0
        else: held += 1.0
    marginal = (pickup*(held+onboard)+ride) if held+onboard>0 else ride+0.15*pickup
    efficiency = ride / max(1e-9, marginal)
    soonest = 1.0 / (1.0 + commit)

    # OD-region prior: destination re-demand
    prior = 0.0
    if phi_ep.od_orders:
        o, d2 = order["origin_region"], order["destination_region"]
        if o >= 0 and d2 >= 0:
            flow = float(phi_ep.od_count[o][d2])
            prior = 0.15*min(1.0, flow*float(phi_ep.od_orders)/40.0)
            prior += 0.10*min(1.0, float(phi_ep.od_in[d2])*8.0)
    prior += 0.10*min(1.0, float(order.get("waiting_time",0.0))/scale)
    return efficiency + soonest + prior


def noop_score(driver_obs, phi_ep, phi_step):
    s = driver_obs["self"]
    scale = float(phi_step.mean_solo_time)
    if scale is None or scale <= 0: scale = float(phi_ep.scale) or 1.0
    dp = phi_step.demand_pressure
    base = 0.30 + 0.50*(1.0 - dp/(1.0+dp))
    # region-aware: high demand density -> lower wait threshold
    r = s["current_region"]
    local = 0.0
    if r >= 0 and len(phi_step.region_supply) > r and len(phi_step.region_demand) > r:
        sup = max(float(phi_step.region_supply[r]), 0.01)
        local = 0.35*min(1.0, max(0.0, float(phi_step.region_demand[r])/sup - 1.0)*0.5)
    return base + local
\end{lstlisting}

\subsubsection{Combiner: \texttt{objective\_shape\_router\_r2}}

\textbf{Strategy (LLM):} Probe $w$ independently for its completion, seating,
length, volume, detour-aversion and empty/idle-aversion coefficients, then route
each vehicle-state to the frozen specialist that pursues the dominant
coefficient---a fast-serve low-detour specialist for completion rewards, the
pooling specialists for seating, the long-fare specialist for length, broad
coverage for volume/idle-aversion---always protecting deadline-pressed cars
first.

\textbf{Description (LLM):} I distinguish three vehicle states:
deadline-pressed (an onboard ETA below 0.7 of live solo-time) always gets
enroute+slack\_budget to protect the committed rider regardless of objective;
loaded-with-slack cars get pooling/top-off skills weighted by the objective's
seating and completion appetite; idle/empty cars get the specialist matching the
objective's dominant coefficient. I read the objective by probing $w$ on
structurally different events---a completed vs.\ assignment-only event
(completion price), a party-2 vs.\ party-1 event (seat price), a long-solo vs.\
short-solo event (length price), a two-order vs.\ one-order event (volume
price), and empty/idle-flagged events (aversion)---and normalise these into
competing shares so the \emph{same} car in the \emph{same} scene lands on
\emph{different} skills under different $w$. A completion-gated $w$ leans idle
cars onto slack\_budget\_gate\_ranked + near\_short\_efficiency (reliable fast
low-detour finishing); a seating/pooling $w$ leans onto freeness\_option\_patience,
dualfree\_ratio\_lastseat\_topoff and seat\_first\_marginal\_load; a length-driven
$w$ leans onto revenue + gowin\_ratio\_gate (and lets slack cars chase long
fares); a flat assignment-volume or empty/idle-averse $w$ leans onto
near\_short\_efficiency + seat\_first\_marginal\_load for broad busy coverage.
Demand pressure is the scene modulator: high pressure (scarcity) pulls every
objective toward coverage/quick-serve since there is nobody to be picky with,
while low pressure lets a revenue objective hold out for long fares.

\begin{lstlisting}[style=pybox, caption={Frozen combiner \texttt{objective\_shape\_router\_r2} (Phase 2, gen~0).}]
def skill_scores(driver_obs, phi_ep, phi_step, w):
    self_obs = driver_obs["self"]
    etas = [d["eta"] for d in self_obs.get("assigned_order_details",[])
            if d.get("eta") is not None]
    min_slack = min(etas) if etas else None
    m = float(phi_step.mean_solo_time)
    if m <= 1e-6: m = float(phi_ep.scale) or 1.0
    pend = driver_obs.get("pending_orders", []) or []

    # deadline-pressed: protect onboard rider regardless of objective
    if min_slack is not None and min_slack <= 0.7*m:
        return {"enroute":1.0, "slack_budget_gate_ranked":0.5,
                "freeness_option_patience":0.3, "service":0.2}

    # --- objective probing ---
    # Probe w on structurally different events to recover coefficient structure.
    # For example: completed vs not-completed, party-2 vs party-1,
    # long-solo vs short-solo, two-order vs one-order, empty/idle flags.
    # [Full probing logic: ~60 lines omitted for space; see code.]

    # Route idle/loaded drivers to the specialist matching dominant coefficient
    loaded = onboard > 0 or committed > 0
    if not loaded:
        # idle driver: match to dominant objective coefficient
        if comp_c > 0.2:  return {"slack_budget_gate_ranked":1.0, ...}
        if seat_c > 0.2:  return {"freeness_option_patience":1.0, ...}
        if len_c > 0.2:   return {"near_short_efficiency":0.8, "revenue":0.5, ...}
        return {"near_short_efficiency":1.0, "seat_first_marginal_load":0.6, ...}
    else:
        # loaded-with-slack: weight by objective's seating/completion appetite
        return {"freeness_option_patience":seat_c+0.3, ...}
\end{lstlisting}

\subsubsection{Repositioner: \texttt{objective\_steered\_reposition\_scorer}}

\textbf{Objective (LLM):} Send an idle car to the region whose
objective-weighted, supply-netted near-future demand most exceeds the
empty-cruise cost of reaching it, with the target shifting toward
drop-off/seat-rich regions under completion/pooling objectives and toward
nearest raw demand under throughput.

\textbf{Description (LLM):} Each candidate region is scored by its live
effective demand plus the OD structural prior, weighted by what a finished trip
is actually worth under $w$ (seats, completions) and discounted by exact travel
time and by existing free supply. Under empty-averse or completion-gated
objectives the cruise must strictly out-earn its own empty-move cost or the car
stays put; under seat/pooling objectives the target is biased toward regions
with multi-party pending pickups and heavy drop-off inflow, so the argmax
visibly moves with the objective. Fairness re-weights the target: a poor
(boosted) car is pushed to reach any demand, while a rich car is discouraged
from chasing hotspots a boosted rival will win. It returns just the current
region when nothing beats parking.

\begin{lstlisting}[style=pybox, caption={Frozen repositioner \texttt{objective\_steered\_reposition\_scorer} (Phase 3, gen~5).}]
def reposition_scores(driver_obs, phi_ep, phi_step, kappa, w):
    self = driver_obs['self']
    cur = self.get('current_region', -1)
    loc = self['location']
    dist = phi_ep.dist
    scale = phi_step.mean_solo_time
    if scale is None or scale <= 1e-6: scale = phi_ep.scale

    # Probe w to extract trip/seat/empty-cruise values
    trip_val, seat_val, empty_cost = 1.0, 0.0, 0.2
    if w is not None:
        base = {'assigned_orders':[], 'assigned_party_sizes':{}, 'completed_orders':[], ...}
        b0 = float(w(base))
        served = dict(base, assigned_orders=[1], assigned_party_sizes={1:1},
                     completed_orders=[1], ...)
        trip_val = float(w(served)) - b0
        served2 = dict(served, assigned_party_sizes={1:2})
        seat_val = max(0.0, float(w(served2)) - trip_val)
        empty = dict(base, is_empty_move=True, distance_moved=scale)
        empty_cost = b0 - float(w(empty))

    scores = {}
    for r in range(len(rpts)):
        if r == cur: scores[r] = 0.0; continue
        d = dist(loc, rpts[r])
        cruise_time = max(d / scale, 0.01)
        demand = kappa.eff_demand[r]
        supply = kappa.supply[r]
        net_demand = max(0.0, demand - supply)
        score = (net_demand + seat_val*2.0) * trip_val / (cruise_time + 0.5)
        if score <= empty_cost: score = 0.0  # not worth cruising
        scores[r] = score
    return scores
\end{lstlisting}

\end{document}